\def\mode{0} 

\if 1\mode
    \documentclass[preprint,superscriptaddress,floatfix]{revtex4-2}
    \usepackage{lineno}
    \linenumbers
    \usepackage{natbib}
\else
    \documentclass[twocolumn,superscriptaddress, aps, prx,floatfix]{revtex4-2}
\fi

\usepackage{graphicx}
\usepackage{dcolumn}
\usepackage{bm}
\usepackage{float}
\usepackage{siunitx}
\usepackage{ amsmath,amssymb}
\usepackage[
	pdftex,
	colorlinks=true,
  	urlcolor=blue,
  	linkcolor=blue,
  	citecolor=blue
]{hyperref}

\renewcommand{\vec}[1]{\ensuremath{\bm{#1}}}

\newcommand{\avg}[1]{\langle #1 \rangle}
\newcommand{\abs}[1]{\ensuremath{\left\vert{#1}\right\vert}}

\newcommand{\ket}[1]{\ensuremath{\left\vert{#1}\right\rangle}}
\newcommand{\bra}[1]{\ensuremath{\left\langle{#1}\right\vert}}

\newcommand{\rhoee}{\ensuremath{\rho_{\mathrm{ee}}}}
\newcommand{\Isat}{\ensuremath{I_{\mathrm{sat}}}}

\begin{document}

\title{Fast single atom imaging for optical lattice arrays }

\author{Lin Su}
\if 0\mode
    \email{lin\_su@g.harvard.edu}
\fi
\affiliation{Department of Physics, Harvard University, Cambridge, Massachusetts 02138, USA}
\author{Alexander Douglas}
\affiliation{Department of Physics, Harvard University, Cambridge, Massachusetts 02138, USA}
\author{Michal Szurek}
\affiliation{Department of Physics, Harvard University, Cambridge, Massachusetts 02138, USA}
\author{Anne H. H\'{e}bert}
\affiliation{Department of Physics, Harvard University, Cambridge, Massachusetts 02138, USA}
\author{Aaron Krahn}
\affiliation{Department of Physics, Harvard University, Cambridge, Massachusetts 02138, USA}
\author{Robin Groth}
\affiliation{Department of Physics, Harvard University, Cambridge, Massachusetts 02138, USA}
\author{Gregory A. Phelps}
\affiliation{Department of Physics, Harvard University, Cambridge, Massachusetts 02138, USA}
\author{Ognjen Markovi\'{c}}
\affiliation{Department of Physics, Harvard University, Cambridge, Massachusetts 02138, USA}

\author{Markus Greiner}
\if 0\mode
    \email{greiner@physics.harvard.edu}
\fi
\affiliation{Department of Physics, Harvard University, Cambridge, Massachusetts 02138, USA}

\date{\today}
\begin{abstract}

High-resolution fluorescence imaging of ultracold atoms and molecules is paramount to performing quantum simulation and computation in optical lattices and tweezers. Imaging durations in these experiments typically range from a millisecond to a second, significantly limiting the cycle time. In this work, we present fast, \qty{2.4}{\us} single-atom imaging in lattices, with 99.4\% fidelity - pushing the readout duration of neutral atom quantum platforms to be close to that of superconducting qubit platforms. Additionally, we thoroughly study the performance of accordion lattices. We also demonstrate number-resolved imaging without parity projection, which will facilitate experiments such as the exploration of high-filling phases in the extended Bose-Hubbard models, multi-band or SU(N) Fermi-Hubbard models, and quantum link models.

\end{abstract}

\maketitle

\section*{Introduction}

Quantum simulation \cite{Jaksch1998, Greiner2002, Gross2021, Altman2021, Daley2022} with ultracold atoms has enabled the generation and exploration of strongly correlated matter. Itinerant models that are challenging to probe with computer simulations have been successfully studied with optical lattice-based quantum simulators\cite{Gross2021, Bohrdt2021, Scholl2021}. Concurrently, atoms in optical tweezer arrays \cite{Gaetan2009, Urban2009, Ebadi2021, Kaufman2021} have ushered in a new era of quantum computation \cite{Graham2022, Ma2023, Scholl2023, Singh2023, Lis2023, Norcia2023} and enabled large-scale quantum algorithms with logical qubits \cite{Bluvstein2024}. Essential to both of these platforms is imaging \cite{McDonald2019, Asteria2021, Milner2023} of individual atoms \cite{Schlosser2001, Nelson2007, Bakr2009, Sherson2010}.

Fast imaging significantly reduces cycle time, which is one of the key figures of merit for quantum simulators and quantum computers. Moreover, quantum error correction, essential to fault-tolerant quantum computation, relies on fast, mid-sequence readout \cite{Bluvstein2024}. Efforts have been made to improve the repetition rate of quantum experiments by greatly accelerating the state preparation of degenerate quantum gases \cite{Phelps2020, Vendeiro2022} or by loading tweezers directly from magneto-optical traps and implementing sideband cooling \cite{Bao2023, Lu2024}. However, high-fidelity imaging takes hundreds of milliseconds in many machines, which can act as an experimental cycle time bottleneck. Cavity-assisted measurements within tens of microseconds have been demonstrated \cite{Deist2022} but require complex optics installed inside vacuum. Recently, fast imaging in \qty{8}{\us} \cite{Su2023} (99.5\% fidelity) and \qty{20}{\us} (98.6\% fidelity) \cite{Ma2023, Scholl2023} have been reported. Yet, neutral atom platforms so far still need much longer readout time than other quantum computing platforms, such as superconducting qubit platforms \cite{Blais2021, Chen2023, Swiadek2023, Sunada2024}.

Another key limitation of imaging systems is their resolution: the diffraction limit of the imaging light can be a significant hurdle to the site-resolved imaging fidelity \cite{Phelps2019}. When studying itinerant models, wavelength-scale lattice spacing is used to optimize tunneling \cite{Gross2021}. Even smaller lattice spacing is required to study long-range interactions \cite{Su2023}, or to realize collective optical effects \cite{Rui2020}. In these cases, the optical point spread function between neighboring sites may substantially overlap, requiring longer imaging times to distinguish between sites with high fidelity. We used accordion lattices to study the extended Hubbard model in our recent work \cite{Su2023} but here we fully characterize the performance of the accordion lattices.

Conventionally, atoms must be localized to the same sites throughout the imaging duration, necessitating deep pinning traps and efficient on-site optical cooling. Imaging multiple atoms per single site leads to parity projection: light-assisted collisions \cite{DePue1999, Schlosser2001, Jones2006, Fuhrmanek2012, Sompet2013} cause pairwise loss of atoms, resulting in the final atom number in a site to be 0 (1) if the initial atom number is even (odd). To avoid this limitation and count the total number of atoms in each site, techniques of expanding the atoms to multiple sites have been demonstrated \cite{Kaufman2016, Preiss2015, Hartke2020, Lebrat2023, Boll2016, Koepsell2020, Prichard2023}. However, direct on-site atom counting would significantly reduce the technical complexity of exploring models with many atoms in a lattice site.

In this work, building upon free space imaging demonstrated in \cite{Bergschneider2018, Su2023}, we push the limit of fast site-resolved imaging. We achieve a high fidelity ($>99.4\%$) of distinguishing between 0 and 1 atoms per site within as little as \qty{2.4}{\us}, bringing neutral atom quantum platforms to be comparable to superconducting platforms in terms of detection time \cite{Swiadek2023, Sunada2024}. We elucidate and overcome the detrimental effects of continuous high-saturation imaging of atoms from two directions and experimentally confirm the understanding of the imaging process. Using an accordion lattice, we resolve a small-spacing lattice beyond the diffraction limit. Moreover, we demonstrate parity-projection-free imaging that for the first time enables direct atom number detection.

\section*{Experimental Setup}

\begin{figure}
    \centering
    \includegraphics[width=0.48\textwidth]{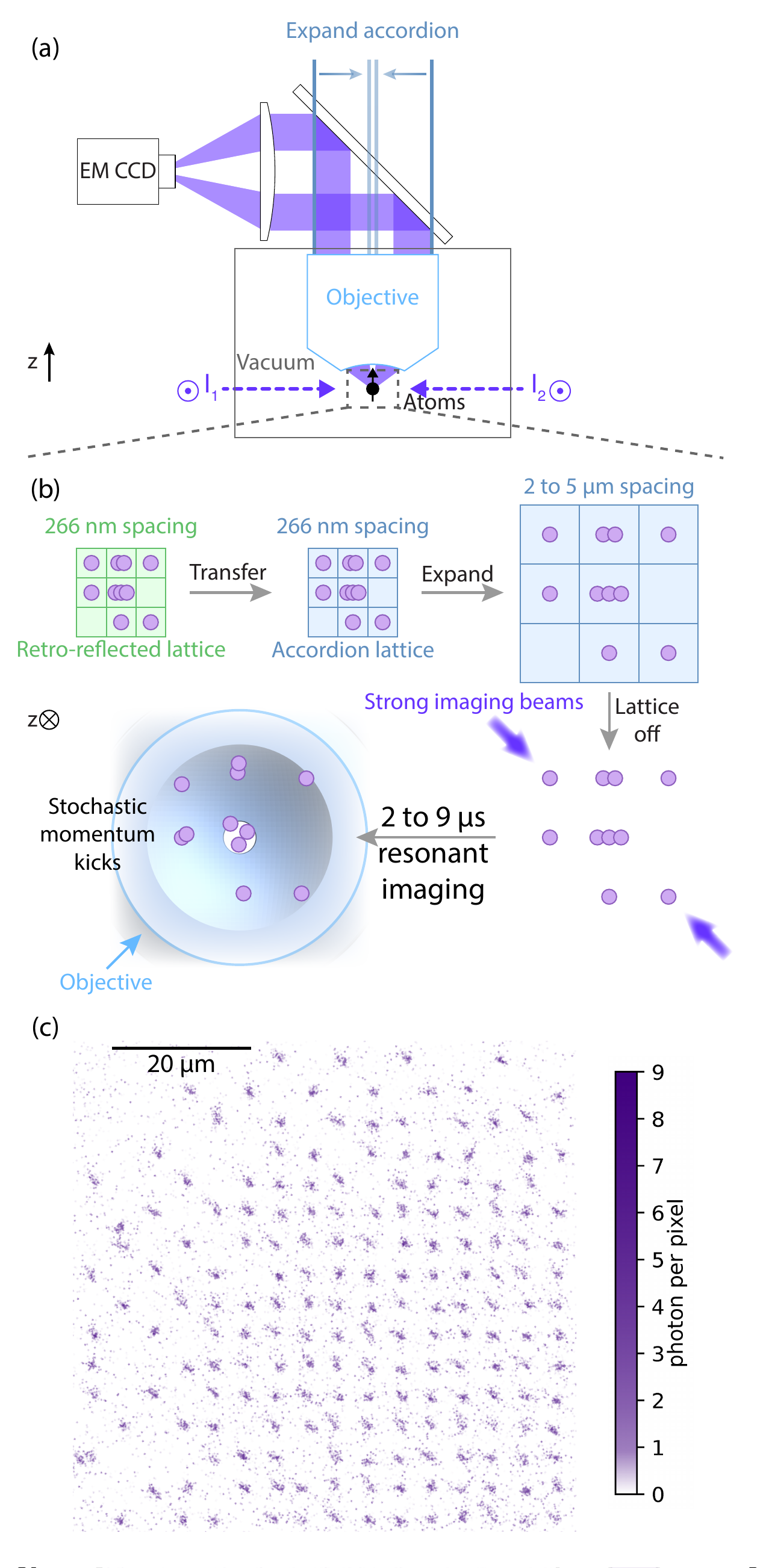}
    \caption{\textbf{Imaging setup.} (a) Our quantum gas microscope consists of a high-resolution objective mounted in vacuum and a set of tunable spacing accordion lattices. To perform imaging, we send two counter-propagating beams onto the atoms and set the beam polarization and atom quantization axis to maximize the objective collection efficiency (SM~\ref{collection_efficiency}). To prepare atoms for free space imaging, we transfer atoms from the tightly spaced lattice (green) to the accordion lattice (blue) and expand the accordion lattice shown in (b). The large spacing allows us to perform fast fluorescence imaging in free space without cooling or trapping and obtain single-shot images. (c) is one such image with a Mott Insulator in the center and shows checkerboard patterns on the edge - a signature of dipolar interactions between the magnetic atoms used in our quantum gas microscope.
    }
    \label{fig: setup}
\end{figure}

The fast imaging method with accordion lattices presented in this work is primarily developed for our quantum gas microscope that uses magnetic erbium atoms to create strongly correlated lattice models with long-range dipolar interaction \cite{Chomaz2022}. This enables quantum simulations of extended Hubbard models and recently led to the observation of dipolar quantum solids \cite{Su2023}. To maximize the magnetic dipolar interaction strength \cite{Du2024}, we choose a lattice spacing of \qty{266}{\nm}, roughly two times smaller than the lattice spacing used in typical experiments. Both resolving this small lattice spacing optically and laser cooling of erbium atoms in lattices during imaging is technically challenging. Therefore, we use accordion lattices \cite{Fallani2005, Huckans2006, Li2008, Al2010} to increase the spacing of the sites during imaging to a similar regime as optical tweezer arrays.  With a spacing of a few microns, the point spread functions of adjacent lattice sites do not overlap. Therefore, we only need to detect as few as 15 photons to identify an atom reliably. We can then completely circumvent trapping and laser cooling during imaging and instead rely on the inertia of the atom to keep it in place as we quickly scatter as many photons as possible \cite{Bucker2009, Bergschneider2018, Ma2023, Scholl2023}. This scheme not only greatly speeds up the imaging to only a few microseconds, but also removes the complexity of pinning lattices and cooling schemes.

In our experiment we rapidly create Bose-Einstein condensates of the bosonic isotope of erbium ($^{168}$Er) within a few hundred milliseconds \cite{Phelps2020}, and adiabatically load these condensates into tightly spaced retro-reflected lattices (green in Figure~\ref{fig: setup}(b)). To image the atoms, we turn off the lattice dynamics by quickly ramping up the tightly spaced lattice and transferring the atoms to an accordion lattice (blue in Figure~\ref{fig: setup}(b)). The accordion lattice is created by projecting beams through a high numerical aperture objective (Figure~\ref{fig: setup}(a) and Supplemental Materials (SM)~\ref{accordion}). The accordion lattice spacing is expanded to a few microns in \qty{80}{ms}. We estimate the loss rate during the lattice transfer and expansion to be smaller than 1\% (SM~\ref{accordion}). After quenching off the lattices, we illuminate the atoms from two sides with high-intensity ($I\approx20I_\text{sat}$, where $I_\text{sat}$ is the saturation intensity of the transition at \qty{56}{\mW/\cm^2}) imaging beams (purple arrows in Figure~\ref{fig: setup}(a)) for a total of a few microseconds on resonance with the broad \qty{30}{\MHz} transition of erbium at \qty{401}{\nm} and scatter 80 photons per microsecond. Fluorescence photons are captured by the objective and recorded on an Electron Multiplying (EM) Charge-Coupled Device (CCD) camera (Figure~\ref{fig: setup}(c)). As a result of a through hole in our custom objective (SM~\ref{objective}), we collect photons between 0.3 and 0.85 Numerical Aperture (NA). It is worth noting that our imaging scheme injects a significant amount of energy into our system and as a result, the atoms are effectively lost after being imaged. However, if trap depths were to be increased to much larger than the kinetic energy imparted during imaging, it is conceivable to recapture atoms after imaging, albeit at a fairly high temperature. 

\section*{Effects of imaging light recoil}

\begin{figure*}
    \centering
    \includegraphics[width=\textwidth]{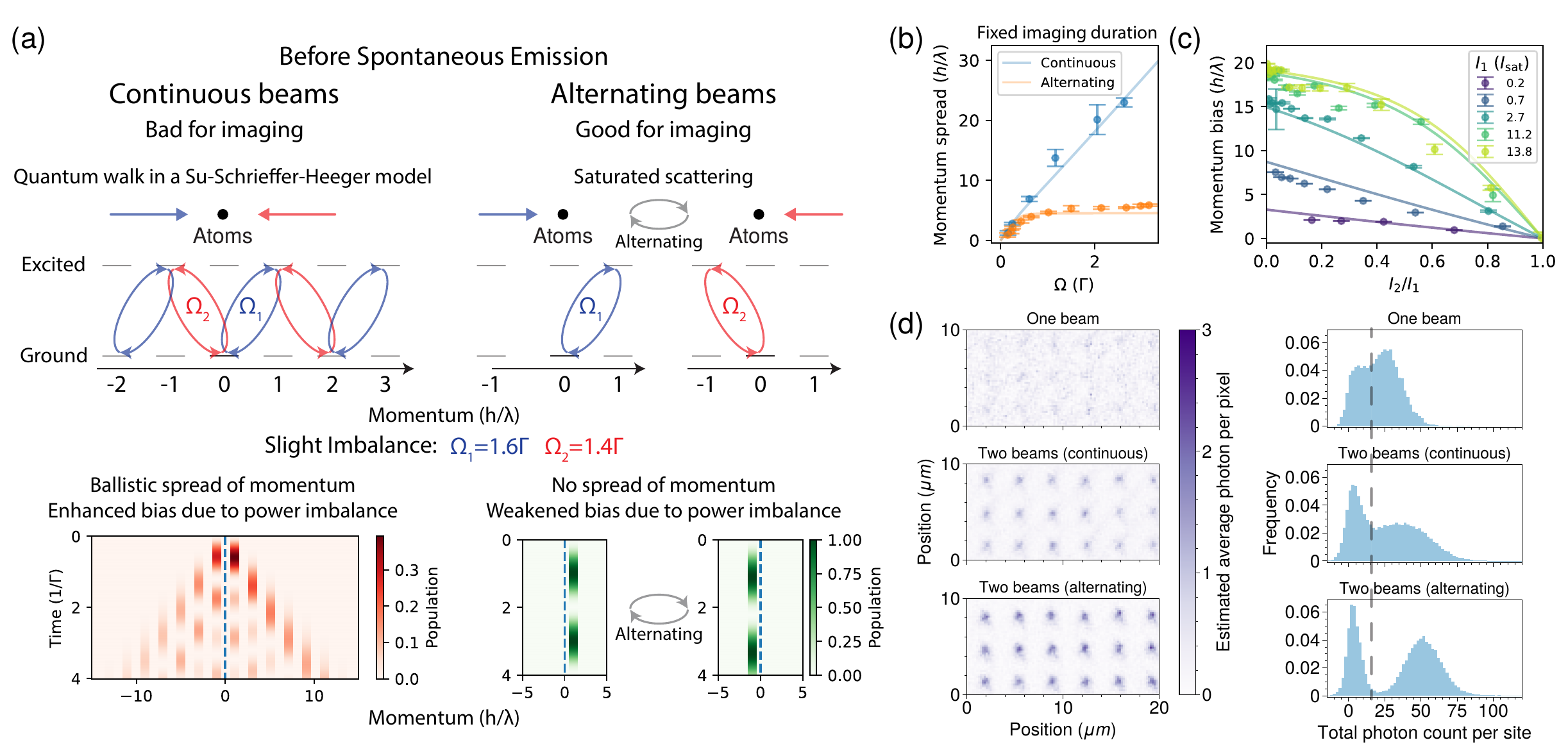}
    \caption{\textbf{Alternating pulsed imaging beams result in high imaging fidelity.} The spread and bias of atom momentum can blur images and hurt imaging fidelity. To prevent atoms from accelerating in a single direction and veering into adjacent lattice sites, we apply two counter-propagating beams to average out the momentum kicks. With simultaneous illumination from two sides with high intensity ($I\approx20\Isat$), however, we observe fast spatial spreading of atoms. Remarkably, this is caused by coherent quantum walks that occur in between spontaneous emission events and that quickly spread out the atomic wave function to span a large number of momentum states. The dynamics for $t\lesssim1/\Gamma$ can be conveniently approximated by coherent oscillation in this highly-saturated regime ($I\gg\Isat$), and this picture can be used to understand interesting features of the dynamics. We simulate the excited state population in the left panel of (a). In addition, the quantum walk enhances the momentum bias due to intensity imbalance when both beams are on simultaneously. With simultaneous beams the momentum during Rabi oscillations spreads out to encompass tens of $\hbar k$ momenta (excited state population shown in red); in contrast, when the two beams are pulsed alternatingly the momentum does not spread beyond $\pm \hbar k$ (excited state population shown in green). In addition, the momentum bias with alternating beams is insensitive to intensity imbalance when $\Omega>\Gamma$ (right panel of (a)). The overall movement during imaging is then largely dominated by the diffusive random walk originating from spontaneous emission events. (b), To take into account spontaneous emissions, we use the master equation solution in SM~\ref{master_equation_simulation}. We consider the momentum spread after many spontaneous emissions. For a fixed imaging duration, with two continuous beams, as the beam intensities are increased, the momentum spread increases linearly without bound, but with alternating beams, the momentum spread no longer increases with higher beam intensities (at a fixed imaging duration). We measure the momentum spread of atoms with time-of-flight images and observe good agreement with the master equation simulation, shown in solid lines. All error bars in this paper are standard error of the mean. To demonstrate the increased sensitivity to the intensity imbalance in the continuous beam configuration, we plot the measured momentum bias (dots with error bars) and simulated results (solid lines) in (c). (d) shows the averaged images of atoms when only one beam is on (top) when both beams are on continuously (middle), and when both beams are alternatingly pulsed on with no overlap (bottom). The total imaging duration is the same at \qty{6.4}{\us}, leading to similar photon flux per atom but the alternating beam image (bottom) is much sharper, qualitatively agreeing with the theory. In addition, the peaks in the histogram are better separated when the two beams are pulsed.
    }
    \label{fig: SSH_chain}
\end{figure*}

\begin{figure}
    \centering
    \includegraphics[width=0.48\textwidth]{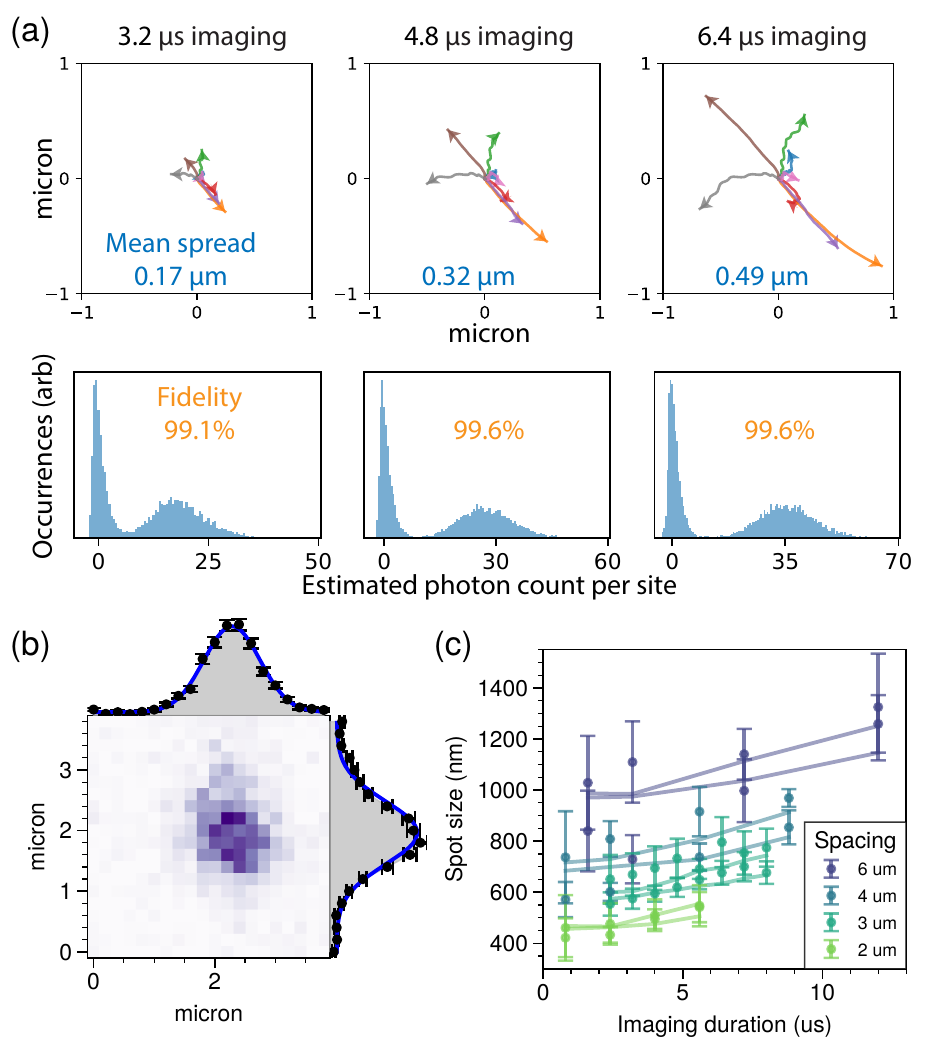}
    \caption{\textbf{Imaging with alternating beams.} With alternatingly pulsed beams, the stochastic recoil causes atoms to spread out when the imaging duration is increased, as simulated in the upper panel of (a), where colors represent different simulated traces. To compare the experiment with the simulation, we measure the spot size of our atoms. (b) is an exemplary image averaged over 30 single-shot images where we expand the accordion lattices to \qty{6}{\um} spacing and perform free space imaging for \qty{7.2}{\us}. The image is slightly asymmetric since the beam is along one direction and our objective has aberrations especially when the atom randomly walks out of the focal plane. The image is then cropped into individual sites and 1D profiles are obtained by summing over the x and y axes. The x and y spot sizes are different since the imaging beams are along only one axis and the objective aberrations are anisotropic (SM~\ref{objective}). The spot sizes with different accordion lattice spacings and imaging durations are shown in (c). The measured spot sizes of the x and y profiles are plotted in circles with error bars. The respective colored lines correspond to the simulated spot size along the x and y directions. The measurements qualitatively agree with the simulation.
    }
    \label{fig:spot_size}
\end{figure}

Our free space imaging method relies on the atom not moving significantly from its original position during imaging. Therefore, we minimize the imaging duration by maximizing the photon scattering rate to collect enough photons before the atom can drift far enough to be mischaracterized as being in a different lattice site. This is achieved by shining highly saturated imaging beams, on-resonant to a broad transition, onto the atom. As there is no cooling and trapping, the atom is accelerated solely by the recoil of the scattering photons. When illuminated by an imaging beam, the atom undergoes a coherent Rabi oscillation between 0 and $\hbar$k units of momentum transfer in the direction of the illuminating light. This continues until spontaneous emission projects the atom, imparting $\hbar$k in the direction of the imaging light and $\hbar$k of momentum in an isotropic random direction. Subsequent scattering events occur in the same manner, with Rabi oscillations between spontaneous emission events, for the duration of imaging. Ideally, to ensure the atom remains spatially localized, we want to balance the directional $\hbar$k momentum kick, that accompanies the isotropic momentum kick \cite{Bergschneider2018}. We use two deeply saturated counter-propagating beams to circumvent the acceleration along the illumination direction to cancel out the directional transport from absorption recoil events. Naively, the saturation should prevent fluctuation in relative power from causing mean field motion. Experimentally, however, we find a strong deviation from this picture. With two radiation fields incident on the atoms at the same time, we observe the atom's momentum during the coherent Rabi drive acts non-trivially; the atom quickly spreads out even if the power of the two beams is carefully balanced. We observe diffusion in momentum that scales with the intensity of the beams, despite the rate of incoherent scattering saturating (Fig 2b). In addition, a slight beam intensity imbalance for deeply saturated beams causes rapid acceleration in one direction (Fig 2c), highlighting how detrimental this effect is in realistic, imperfect experimental systems. 
We observed similar phenomena for a range of relative detunings of the counter-propagating beams and different polarization arrangements. Such behavior was previously reported\cite{Bergschneider2018}, but no explanation was found. Here we take careful consideration of the physics of imaging with counter-propagating beams, in particular, what occurs during the Rabi drive in the time between spontaneous emission events, and explain why our solution of using alternating beams circumvents unwanted transport.

To study this surprising effect, we consider a picture where the atom performs a quantum walk in momentum space \cite{Meier2016, Ryu2006}. In between spontaneous emission events, when both beams are on at the same time and couple the atom to the same $m_J$ state with Rabi frequencies $\Omega_1$ and $\Omega_2$, the atom goes through quantum walks in momentum space, which can be described as a realization of the Su-Schrieffer-Heeger (SSH) model \cite{Su1979, Atala2013, Meier2018, Leseleuc2019} (SM~\ref{SSH_mapping}). We simulate this model using exact diagonalization \cite{Weinberg2019}  (Figure~\ref{fig: SSH_chain}(a) left, SM~\ref{imbalanced_SSH}). With high beam intensity, the Rabi frequency from the imaging beams ($\Omega$), which corresponds to two times the tunneling energy in the SSH model, is larger than the decay rate ($\Gamma$). Therefore, the quantum walk spreads the atom wave function to momentum states larger than $h/\lambda$, where $\lambda$ is the imaging wavelength at \qty{401}{\nm}, before spontaneous emission projects the atom's momentum state. In contrast, by alternatingly pulsing the two beams, the atom coherently oscillates between $0$ and $\pm h/\lambda$ momentum states and does not venture into states with higher momentum until spontaneous emission, limiting the momentum spread per scattered photon. Intuitively, with both beams on simultaneously, the atom can act as a conduit to shuttle photons between the two imaging beams and hence takes on a wide spread of momentum to do so; however, with alternating beams only one radiation field has an effect on the coherent drive in a given instance. This not only allows the momentum spread to be narrower but also makes the momentum imbalance very insensitive to beam intensity imbalance, as long as both beam intensities are much higher than $\Isat$. Taking into account the spontaneous emission, we arrive at similar results (SM~\ref{master_equation_simulation}).

To measure the momentum spread in the two imaging configurations, we turn on the beam(s) for a short duration of \qty{0.2}{\us} and then take an image after \qty{100}{\us} time-of-flight. We fit the image and plot the width of the Gaussian fit (Figure~\ref{fig: SSH_chain}(b)). When both beams are on at the same time, the width (blue) increases roughly linearly with the Rabi frequency, which is consistent with the simulation (SM~\ref{master_equation_simulation}). In contrast, when only one beam is on, as in the case of alternatingly pulsed imaging beams, the width (orange) saturates. The spread of momentum states increases the spread of the atom position, directly reducing the imaging fidelity. In addition, we study the enhanced sensitivity at high intensity with continuous beams. Due to the quantum walk interference, any beam intensity imbalance can cause a large bias to the momentum, quickly accelerating the atom along the stronger beam direction and greatly affecting the imaging fidelity (Figure~\ref{fig: SSH_chain}(c) and SM~\ref{light_force}). As shown in Figure~\ref{fig: SSH_chain}(d) left, the peak on the middle panel is significantly lower than the one on the bottom panel, since atoms spread out much faster and venture into neighboring sites, resulting in worse fidelity (Figure~\ref{fig: SSH_chain}(c) right). Therefore, we pulse the beams alternatingly so that as long as each beam is highly saturated and the excited state population is close to 50\%, the recoil is relatively insensitive to fluctuations in the beam power. 

Our measurements and theory explain why alternating imaging beams give significantly higher fidelity. Hence, from here on, we optimize the imaging in this configuration. We perform a simulation of the atom's trajectory during imaging and of the fluorescence signal. The simulation takes into account the stochastic recoil from imaging light on the atoms inside and out of the focal plane and other factors (SM~\ref{simulation}). As atoms are imaged for a longer duration, they spread out further around the original location (Figure~\ref{fig:spot_size}(a) top). Meanwhile, more photons are collected so the histograms are separated further, increasing the estimated fidelity (Figure~\ref{fig:spot_size}(a) bottom), ultimately limited by the transition to dark states as discussed in the next section. To confirm the accuracy of our simulation, we compare the average image between the simulation and the experimental data. The accordion lattices are set to different spacings and the imaging beams are set to different durations. We then fit the average image with a Gaussian function and define the spot size as twice the standard deviation (Figure~\ref{fig:spot_size}(b)). The imaging spot size increases for larger accordion spacing because of slight shot-to-shot position fluctuations of the accordion lattice due to air currents (0.065 waves standard deviation) \cite{Phelps2019}. In addition, the calculated ground state Wannier functions based on the measured lattice depth and spacing of the accordion lattices also grow larger when the spacing is larger. Furthermore, the simulation takes into account the wavefront error of the objective that was measured before its placement in the vacuum chamber \cite{Krahn2021}. The data agrees with the simulation as shown in Figure~\ref{fig:spot_size}(c), confirming our understanding of the imaging process and facilitating the application of this method to different atoms or molecules.

\section*{Ultra-fast imaging using binarization}

\begin{figure*}
    \centering
    \includegraphics[width=\textwidth]{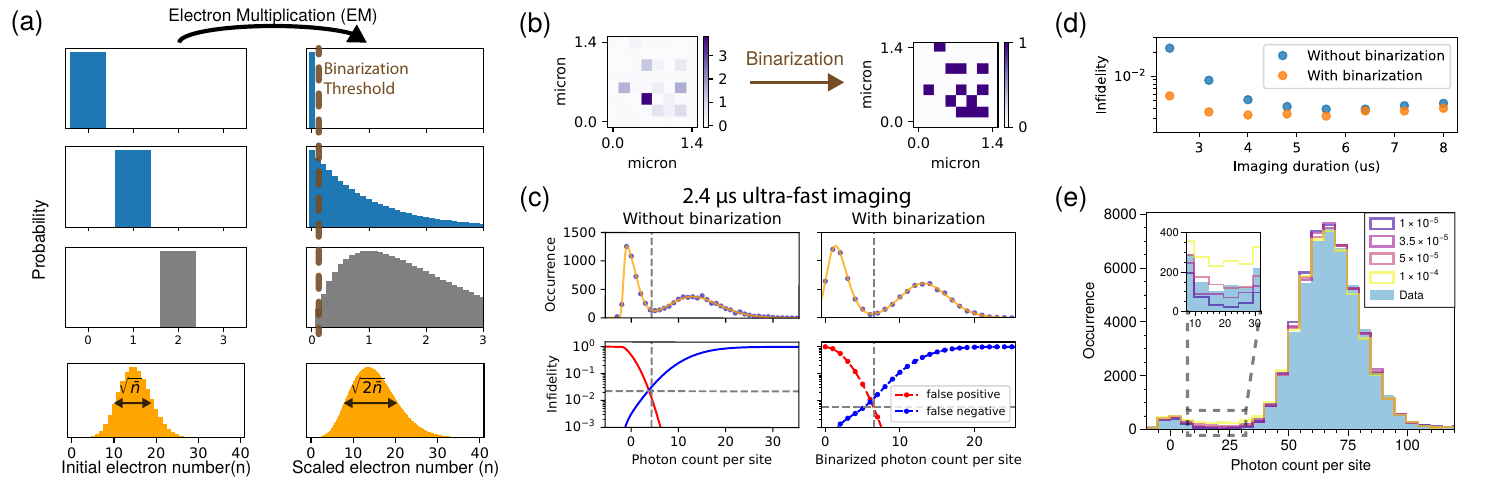}
    \caption{\textbf{Ultra-fast high-fidelity imaging via binarization of Electron Multiplying (EM) CCD counts.} EM noise on the EM CCD camera can negatively affect the signal-to-noise ratio (a). Pixels with 0, 1, or 2 initial electrons (left top 3 panels) are amplified via EM, resulting in overlapping probability distributions that make it impossible to precisely distinguish the initial electron number (right top 3 panels). Assuming a Poisson distribution of the initial electron number (left lowest panel in orange), the EM process results in a factor of two more variance (right lowest panel in orange). When the photon density per pixel is less than one (the blue panels dominate), however, setting a binarization threshold (brown dashed line) to each camera pixel enables a near-perfect distinction between 0 and 1 initial electrons. This effectively removes the EM noise and increases the signal-to-noise ratio. An example of binarization is shown in (b), where only 15 photons are collected on the camera within \qty{2.4}{\us} (\qty{3}{\um} accordion lattice spacing). The histograms are fitted with two skew-normal distributions and a constant offset between the peaks to account for the branching ratio in (c). The infidelity at different cutoffs is estimated based on the fit, assuming the overlapping distributions are accurate representations of the probability distribution in the tails. Binarization increases the estimated fidelity from 97.7\% to 99.4\% on our EM CCD camera. The maximum fidelity of more than 99.5\% can be achieved with only \qty{3}{\us} imaging duration as shown in (d). The fidelity is mainly limited by the atom transitioning into a ``dark state" during imaging, which we estimate in (e) by preparing a cloud with mostly one atom per site and then performing imaging for \qty{8.8}{\us} at \qty{4}{\um} accordion lattice spacing. In addition to two peaks corresponding to 0 and 1 atom per site, we identify significant counts between the peaks with this long imaging duration. The simulated histograms assuming different branching ratios are laid on top of the data, showing an estimated branching ratio of slightly below $5\times10^{-5}$. 
    }
    \label{fig:fig_Branching_Ratio_and_Binarization}
\end{figure*}

The speed of imaging with our EM CCD camera can be further improved using an image processing technique: binarization \cite{Bergschneider2018}. We present a detailed analysis of this technique and use the understanding to reach imaging with 99.4\% fidelity in only \qty{2.4}{\us} after collecting 15 photons.

The number of photons collected by the imaging system follows a Poisson distribution (Figure~\ref{fig:fig_Branching_Ratio_and_Binarization}(a) bottom left). Standard practice for fluorescence imaging techniques is to determine optimal cutoffs between different modeled photon count distributions to identify the presence or absence of an atom. In an ideal system, this is done by choosing the cutoff at the minimal overlap between two Poissonian distributions. However, EM CCD cameras introduce additional noise during the EM process and broaden the distribution to twice the original variance (Figure~\ref{fig:fig_Branching_Ratio_and_Binarization}(a) bottom right) so careful consideration of how to model this noise is necessary to faithfully reconstruct the probability distributions. The imaging magnification is chosen so that each pixel mostly detects only zero or one photon (Figure~\ref{fig:fig_Branching_Ratio_and_Binarization}(a) blue histograms). Applying a binarization threshold (brown dashed line) to each pixel can effectively distinguish whether the pixel detected a photon or not and strongly suppress the additional EM noise when summing up pixels to obtain histograms \cite{Krahn2021} (Figure~\ref{fig:fig_Branching_Ratio_and_Binarization}(b)). Fundamentally, this cut-off method is a post-processing technique that improves signal-to-noise ratio, and hence readout accuracy, by taking advantage of how nonlinear electron multiplication noise factors in differently when distinguishing between 0 and 1 photon distributions and between 1 and 2 (or any higher number) photon distributions. We compare the histograms with and without applying binarization when the atoms are imaged for only \qty{2.4}{\us}. In Figure~\ref{fig:fig_Branching_Ratio_and_Binarization}(c), we fit the histograms with two skew-normal distributions (since the photon number is low, the Poisson distribution due to photoelectron number and the Gamma distribution due to electron multiplication are both asymmetric) and add a constant background in between the Gaussian peaks to account for erbium's branching ratio. The infidelity of false positive (red in Figure~\ref{fig:fig_Branching_Ratio_and_Binarization}(c)) and false negative (blue) errors at different cutoffs are shown. For the case without binarization, we plot the distribution continuously. For the case with binarization, we plot the discrete distribution. Based on the cumulative distributions, we use gray dashed lines to mark the total count cutoff that gives the smallest sum of infidelity for false positives and false negatives. Binarization reduces the variance of the peaks by roughly 1.8 times and increases the fidelity significantly from 97.7\% to 99.4\%. This improvement of signal-to-noise ratio can be exploited when fast imaging with few photons is desired, such as in tweezer array experiments \cite{Ma2023, Bluvstein2024, Scholl2023, Graham2022, Singh2023, Lis2023, Norcia2023}.

Our fidelity at long imaging duration (Figure~\ref{fig:fig_Branching_Ratio_and_Binarization}(d)) is limited by the atom transitioning into intermediate long-lived ``dark" states. Once an atom decays to a dark state, it no longer scatters photons and can be mischaracterized as absent. This chance of decay to dark states (the branching ratio) can be estimated by analyzing the histogram of the counts in between the peaks corresponding to 0 and 1 atoms per site. We simulate the histogram taking into account the stochastic random walk of atoms in and out of the focal plane as well as an assumed branching ratio. The result indicates a branching ratio of slightly less than $5\times10^{-5}$. Similar results are obtained with different imaging durations and accordion lattice spacing. Previous calculations estimated the total branching ratio to dark states to be on the order of $10^{-4}$\cite{Ban2005} and measurements indicate the ratio to be slightly less than $10^{-5}$ \cite{McClelland2006}, which is roughly consistent with our estimate. Adding repumper lasers or choosing a transition with a lower branching ratio could further improve imaging fidelity.

\section*{Parity-projection-free imaging}

\begin{figure}
    \centering
    \includegraphics[width=0.48\textwidth]{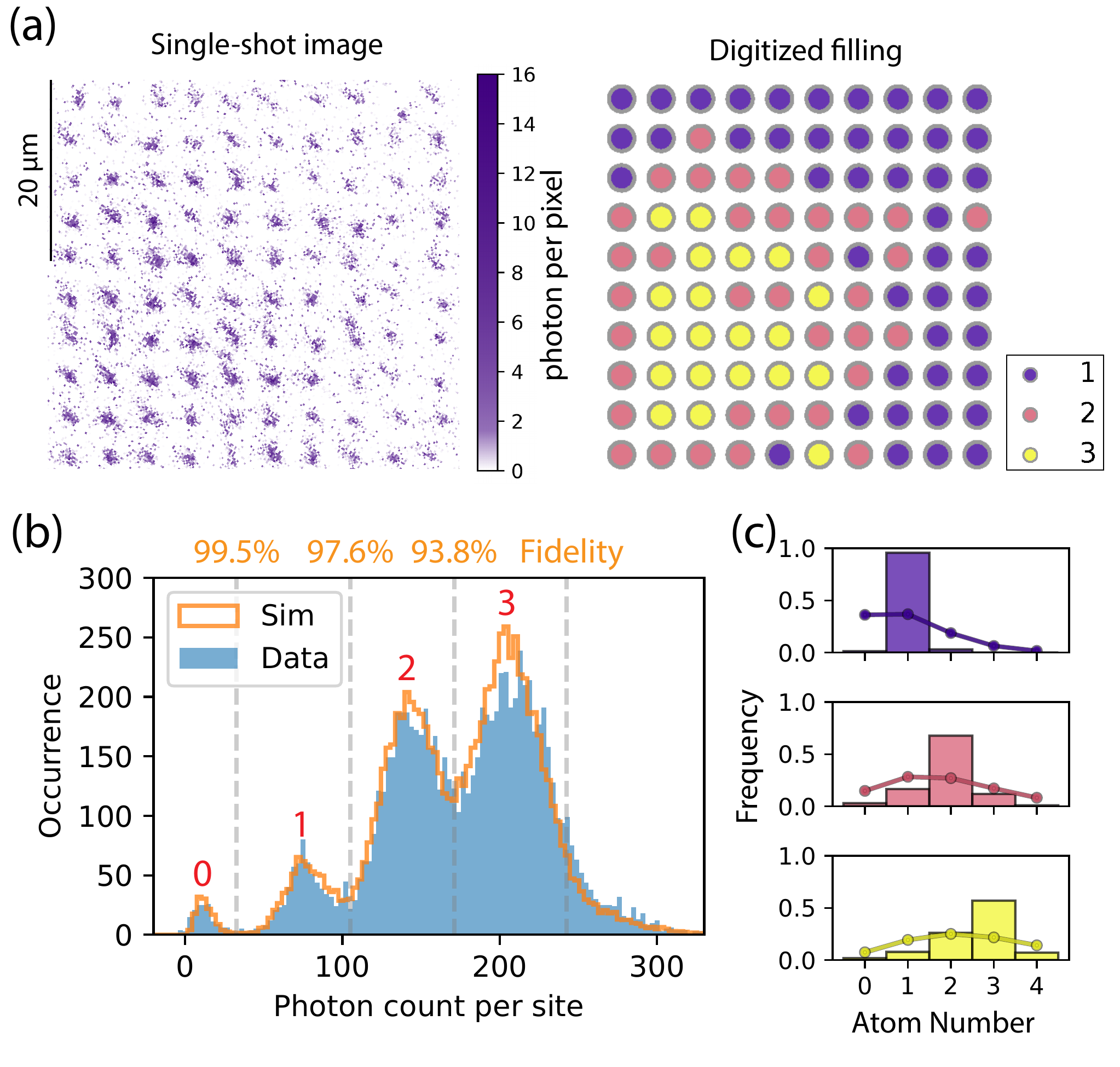}
    \caption{\textbf{Parity-projection-free imaging.} A single-shot image of more than one particle per site is shown in (a). The histogram of the data and the simulation (no free parameters except the peak heights) shows good agreement in (b). The fidelity from the simulation is labeled on top of the graph. With 700 digitized single-shot images, we select three sites dominated by 1, 2, and 3 atoms filling and plot the histogram of the atom number in (c). The Poisson distribution is overlaid in solid lines to contrast the sub-Poissonian statistics we observed, qualitatively showing that we are in the Mott Insulator regime.
    }
    \label{fig:multiple_hist}
\end{figure}

This imaging method uniquely enables direct atom number readout without parity projection for the first time. We load multiple atoms per site and measure histograms by summing the raw camera count in a $19\times19$-pixel box (Figure~\ref{fig:multiple_hist}(b) light blue bars), which shows distinct peaks, allowing different atom numbers to be distinguished. Our simulation (SM~\ref{simulation}) with only the peak heights as free parameters is plotted in orange, where the simulated photon count peak centers corresponding to different numbers of atoms per site match the measured results. The peak shapes are also similar between simulation and experiment. In the simulation, we assume no interactions between atoms on the same site, so the close match between measurement and simulation is evidence that we do not suffer from parity projection due to light-assisted collisions. The simulated fidelity is labeled on top of Figure~\ref{fig:multiple_hist}(b). The demonstrated parity-projection-free imaging (Figure~\ref{fig:multiple_hist}(a) right) has enabled us to observe the sub-Poissonian statistic of the Mott insulator \cite{Folling2006, Bakr2010, Sherson2010, Preiss2015} with more than one particle per site, in situ, in Figure~\ref{fig:multiple_hist}(c). The fidelity could be further improved by employing the latest camera technology and expanding the accordion lattice further.  With qCMOS cameras free from EM noise and proper magnification, we estimate roughly 99\% fidelity distinguishing between 1 and 2 atoms per site and more than 95\% fidelity distinguishing between 3 and 4 atoms per site. By increasing the dynamic range of the accordion lattice, we would be able to space the atoms further than \qty{4.5}{\um} and image longer to resolve even larger atom numbers.

\section*{Conclusion and outlook}

We have thoroughly studied and optimized the imaging method that features an imaging duration of only \qty{2.4}{\us}, pushing the state-of-the-art readout time of neutral atom to one step closer to that of superconducting qubits \cite{Swiadek2023, Sunada2024}. We demonstrated high fidelity of 99.5\% between 0 and 1 atoms per site and more than 90\% up to 4 atoms per site. Cameras with lower noise and the application of machine learning \cite{Picard2020, Impertro2023, Phuttitarn2023} may further increase the performance of our imaging method. Our technique can be applied to any atom or molecule in optical tweezers or lattices and requires only a few hundred photons scattered, which can be essential for the detection of molecules whose transitions are not fully closed \cite{Cornish2024, Langen2024}. For tweezers, the already large spacing eliminates the need for accordion lattices, reducing the total imaging duration from the typical value of a few milliseconds to only a few microseconds. Although the survival probability of atoms after imaging in the trap is small, many quantum error correction methods with mid-circuit readout do not require the original imaged atom to be preserved \cite{Ma2023, Scholl2023}. For small-spacing lattices, the accordion lattice transfer and expansion limit the total imaging duration; however, the duration is still favorable compared to the typical imaging duration of quantum gas microscopes using established techniques that are on the order of half a second. Hence, for quantum simulators that do not require an accordion lattice, our fast-imaging technique can be a marked improvement. Moreover, the accordion operation time could still be further reduced to a few milliseconds by increasing lattice depths.

Moreover, we demonstrated that our site-resolved imaging does not suffer from parity projection, which simplifies the technical challenges of studying a wide range of physics, including the entanglement entropy measurement in 2D to study quantum phase transitions and quantum critical points \cite{Kaufman2016, Rispoli2019}, multi-band Fermi Hubbard models \cite{Dopf1990}, SU(N) physics \cite{Hofrichter2016, Taie2022, Lee2018, Pasqualetti2024}, charge density wave and topological phases like Haldane Insulator \cite{Torre2006, Berg2008, Lacki2024} with dipolar atoms or molecules, and quantum link model simulations \cite{Osborne2023}.

\paragraph*{Data and code availability}
The experimental data and simulation code supporting this study's findings are available from the corresponding authors upon reasonable request.

\paragraph*{Acknowledgements}
We are grateful for the early contributions to building the experiment from S. F. Ozturk, S. Ebadi, S. Dickerson, and F. Ferlaino. We acknowledge fruitful discussions with P. Preiss, M. Lebrat, Y. Li, J. Lyu, Y. Lu, E. J. Davis, B. Bakkali-Hassani, A. Kale, and L. Kendrick. We are supported by U.S. Department of Energy Quantum Systems Accelerator DE-AC02-05CH11231, National Science Foundation Center for Ultracold Atoms PHY-1734011, Army Research Office Defense University Research Instrumentation Program W911NF2010104, Office of Naval Research Vannevar Bush Faculty Fellowship N00014-18-1-2863, Gordon and Betty Moore Foundation Grant GBMF11521, and Defense Advanced Research Projects Agency Optimization with Noisy Intermediate-Scale Quantum devices W911NF-20-1-0021. A.D. acknowledges support from the NSF Graduate Research Fellowship Program (grant DGE2140743).

\paragraph*{Author contributions}
L.S., O.M., A.H.H., A.K, A.D., M.S., G.A.P., and R.G. contributed to building the experiment set-up. L.S., A.D., and M.S. acquired the data in the experiment. L.S. analyzed the data. L.S. and O.M. performed the simulation. L.S., O.M., M.S., and A.D. contributed to the manuscript. All authors discussed the results. M.G. supervised all work.

\paragraph*{Competing interests}
M.G. is a cofounder and shareholder of QuEra Computing. All other authors declare no competing interests.

\newpage

\section*{Supplemental Materials}

\subsection{Light recoil}

We extensively discuss atom recoil due to imaging light. First, we consider the SSH model assuming no spontaneous emission, which is more familiar to readers in the field of many-body physics: we clarify the Hamiltonian mapping to the SSH model in ~\ref{SSH_mapping} and then extend the simulation in the main text to spatially localized atoms in ~\ref{imbalanced_SSH}. Second, we include the effects of spontaneous emission and use the master equation to simulate the momentum distribution in ~\ref{master_equation_simulation}, assuming spatially delocalized atoms. Third, we analytically derive the expectation value of the photon recoil in ~\ref{light_force}, which offers a convenient formula to estimate the mean photon recoil without having to go through numerical simulations. Lastly, we discuss the experimental implications in ~\ref{experimental_implications}.

\subsubsection{SSH model Hamiltonian mapping}
\label{SSH_mapping}

The SSH model in its original form has the following Hamiltonian:

$$H=t_1\sum_n\ket{n,B}\bra{n,A}+t_2\sum_n\ket{n+1,A}\bra{n,B}+\mathrm{h.c.}$$

where $t_1$ is the tunneling energy to hop within a unit cell, $t_2$ is the tunneling energy to hop between two unit cells, and $n$ is the integer label for unit cells \cite{Batra2020}.

In our case without spontaneous emission, the excited (e) and ground (g) states can be viewed as A and B sub-lattices in a 1D lattice chain. Our Hamiltonian is:

$$H=\frac{\Omega_1}{2}\sum_p\ket{p,g}\bra{p+1,e}+\frac{\Omega_2}{2}\sum_p\ket{p-1,e}\bra{p,g}+\mathrm{h.c.}$$

where $\Omega_1$ ($\Omega_2$) is the Rabi frequency of the beam coming from the left (right) shown in blue (red) as shown in Figure~\ref{fig: SSH_chain}(a) and $p$ is the integer label for the momentum.

The energy of different momentum states differ slightly due to the kinetic energies, but the recoil energy is four orders of magnitude smaller than the linewidth of the transition. With a few hundred photons scattered, the kinetic energy is still small compared to the linewidth of the transition, so we neglect this effect.

\subsubsection{Momentum kicks on spatially localized atoms with imbalanced Rabi frequencies}
\label{imbalanced_SSH}

\begin{figure}
    \centering
    \includegraphics[width=0.48\textwidth]{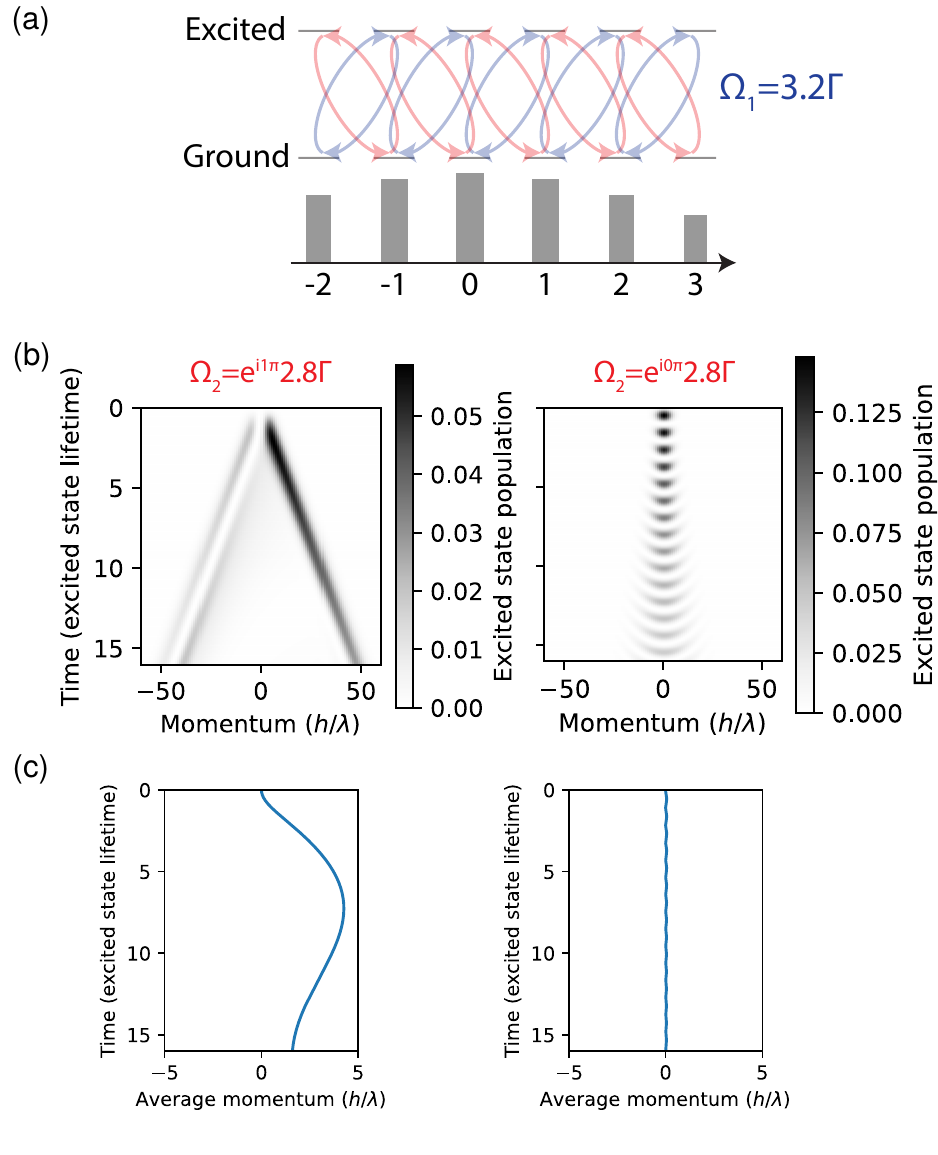}
    \caption{\textbf{Spatial dependence of momentum kicks with continuous non-alternating imaging beams.} When the atomic wave function is localized in a region smaller than the imaging wavelength, the momentum space initial wave function spans a few states as shown in (a). The initial Gaussian state evolves differently depending on the relative phase between the Rabi oscillations from the two beams at specific locations, the initial Gaussian state evolves differently. In the case where the two Rabi frequencies are $\pi$ out of phase, the quantum walk quickly spreads the wave function to very large momentum states as shown in the left panel of (b), with a strong bias in the average momentum as shown in the left panel of (c). When the two Rabi frequencies are in phase, the initial Gaussian wave function does not spread much as shown in the right panel of (b), and the average momentum kick is near zero as shown in the right panel of (c).
    }
    \label{fig: Gaussian_SSH}
\end{figure}

In the main text Figure~\ref{fig: SSH_chain}, we discussed the momentum kicks on atoms assuming the initial atom momentum is a delta function when two beams are on simultaneously. Therefore, the atom wave function spreads more than the wavelength of the imaging beam in real space, washing out any sub-wavelength spatial structure. Here, we relax this assumption and consider realistic wave function size to simulate the spatial dependence of the momentum kick. Based on the measured lattice depth and spacing, we compute the initial momentum state populations based on the Wannier functions (Figure~\ref{fig: Gaussian_SSH}(a)). Different real space locations of the atom give rise to different relative phases in the Rabi frequencies from the two beams. We show two examples when the phase difference is $0$ and $\pi$ to illustrate the spatial dependence of the momentum kicks. Moreover, a small imbalance in the beam intensities results in a large momentum state imbalance. The average momentum for the $\pi$ phase difference is much larger than $h/\lambda$ since the atom absorbs photons from one beam and performs coherent stimulated emission into the other beam before undergoing spontaneous emission. This idea might be developed further to potentially slow atoms or molecules that are unable to spontaneously emit a significant of photons before decaying out of a cycling transition \cite{Nolle1996, Goepfert1997, Shuman2009, Kozyryev2016, Long2019, Kozyryev2018, Galica2018}.

\subsubsection{Simulation of momentum population with spontaneous emission using master equation}
\label{master_equation_simulation}

\begin{figure}
    \centering
    \includegraphics[width=0.48\textwidth]{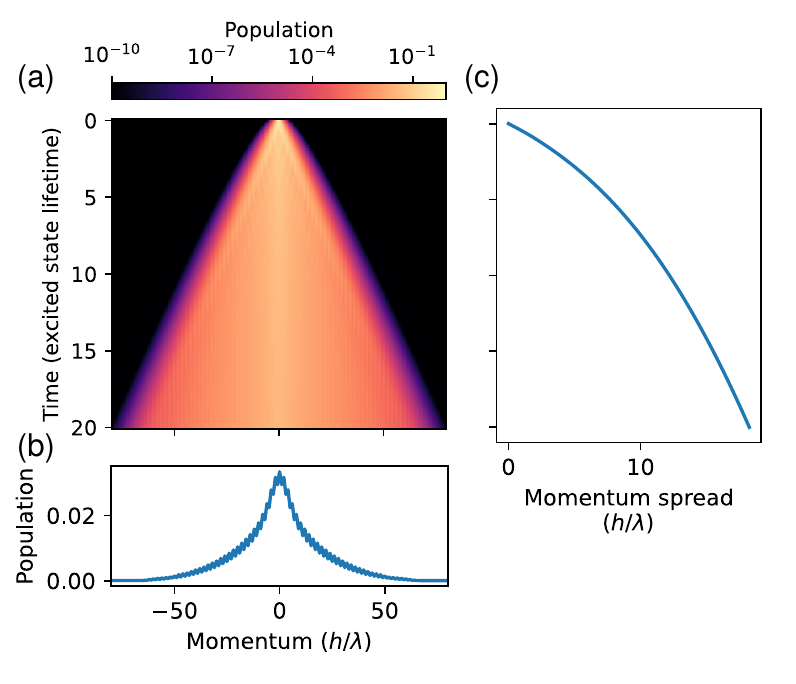}
    \caption{\textbf{Diffusion of momentum with spontaneous emission.} The momentum state of an atom with Rabi drive and spontaneous emission can be simulated by solving the master equation. In (a), we see a light-cone-like boundary in momentum space due to the quantum walk before each spontaneous emission. However, since each spontaneous emission destroys an atom's coherence, the overall momentum evolution is not ballistic. Hence, the population of momentum states remains peaked at 0 as time evolves (b), and the standard deviation of the momentum population increases slower than linearly relative to the evolution time (c).
    }
    \label{fig: master_equation}
\end{figure}

The SSH model offers intuition into the ballistic quantum walk of the atom's momentum before spontaneous emission. To take into account spontaneous emission, we solve the master equation \cite{Campaioli2023} using QuTiP \cite{Johansson2013} and show the momentum state population evolution in Figure~\ref{fig: master_equation}(a). Like in the main text but unlike in SM~\ref{imbalanced_SSH}, we assume the initial state is fully localized at the zero momentum state for simplicity. The momentum population distribution after an evolution of 20 excited state lifetimes is shown in Figure~\ref{fig: master_equation}(b). After each spontaneous emission, the momentum state decoheres, resulting in an overall diffusion of momentum, as shown by the nonlinear shape of the momentum spread in Figure~\ref{fig: master_equation}(c). Although the shape is not a perfect parabola, the non-linear shape shows a clear distinction from the light cone shape in ballistic propagation. The diffusion of momentum versus time is not to be confused with the almost linear relation between momentum spread and Rabi frequency of the imaging light discussed in the main text, which comes from the ballistic quantum walk before each spontaneous emission when $I\gg\Isat$.

\subsubsection{Analytical derivation of the average photon recoil}
\label{light_force}

\begin{figure*}
    \centering
    \includegraphics[width=\textwidth]{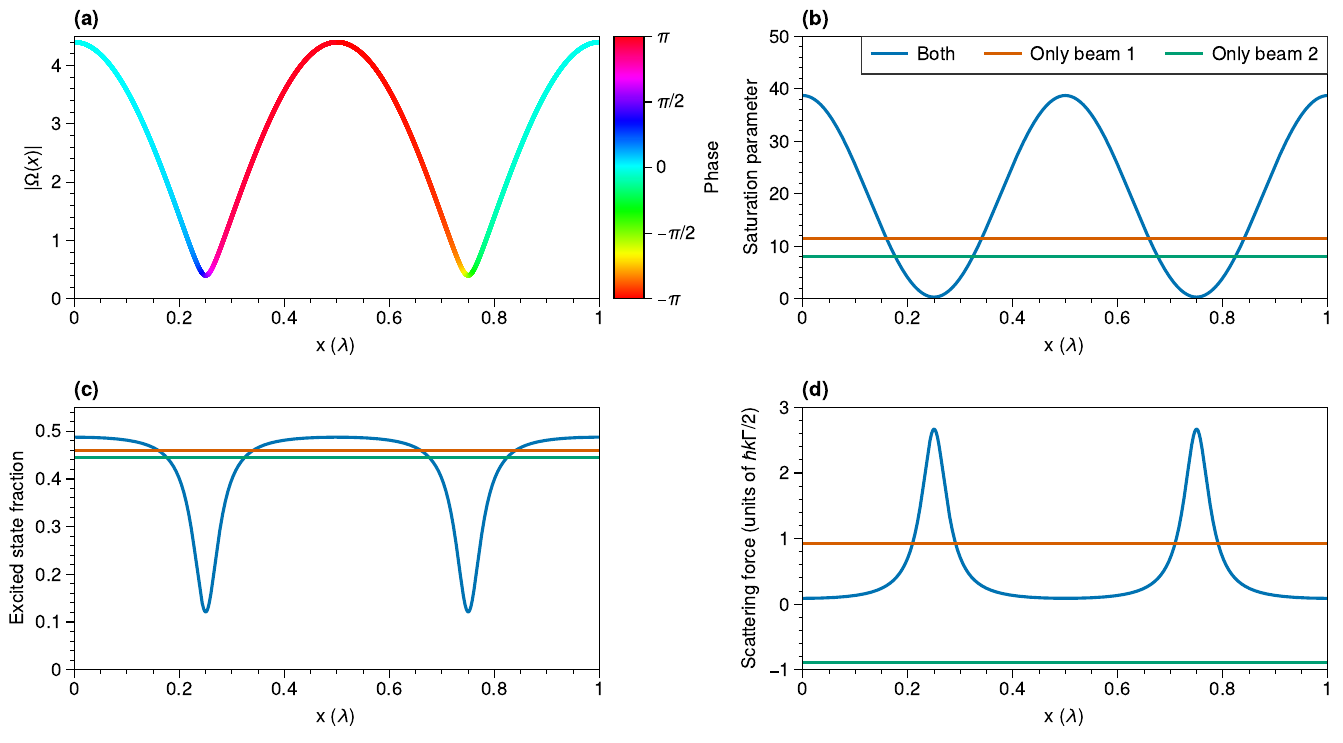}
    \caption{\textbf{Spatial radiation pressure force.} Here we simulate two counter-propagating beams with intensity imbalance, where $I_1=11.52I_\text{sat}$ and $I_2=8I_\text{sat}$. The beams form a standing wave as shown in (a), resulting in the spatial variation of the saturation parameter as shown in (b). We also plot the case when only beam 1 or 2 is on for reference. The excited state fraction is shown in (c). The atoms experience a strong mean radiation pressure force at the minimum of the standing wave biased towards one direction as shown in (d).
    }
    \label{fig:spatial_scattering_force}
\end{figure*}

We derive the average momentum exerted by light beams on a two-level atom analytically to study the dependence of momentum kicks on the beam intensity imbalance. The results agree with the master equation approach described in SM~\ref{master_equation_simulation}.

This semi-classical approach is outlined in Ref.~\cite{steck2007quantum}, Section~5.8.4, which is based on Ref.~\cite{PhysRevA.21.1606}. In this picture, the atoms experience a force induced by a position-dependent light amplitude. We consider a two-level atomic dipole and classical electric fields.

For a general light field, its amplitude can be represented by a spatially varying Rabi frequency
\begin{equation}
        \Omega(\vec{r}) = \abs{\Omega(\vec{r})} e^{i \phi(\vec{r})},
\end{equation}
where $\phi(\vec{r})$ describes the spatially varying phase of the light field.

Assuming the atom moves slowly on timescales of the excited state decay time, the mean light force on the atom is \cite{steck2007quantum}
\begin{equation}
    \avg{\vec{F}} = \frac{\hbar s(\vec{r})}{1 + s(\vec{r})} \bigg ( - \Delta \nabla \log \abs{\Omega(\vec{r})} + \frac{\Gamma}{2} \nabla \phi(\vec{r})\bigg ).
\end{equation}
The spatially dependent saturation parameter $s(\vec{r})$ is
\begin{equation}
    s(\vec{r}) = \frac{\abs{\Omega(\vec{r})}^2}{2 [(\Gamma/2)^2+\Delta^2]},
\end{equation}
where $\Delta$ is the detuning of the light from resonance and $\Gamma$ is the excited state decay rate.

The two terms in the expression for the mean force can be identified as the dipole force and the radiation pressure force, respectively. In this work, we work on resonance ($\Delta=0$), such that the dipole force is zero and the only contributing force is the radiation pressure force. The radiation pressure force can be rewritten by noting that $s(\vec{r})/[2(1+s(\vec{r}))]=\rhoee(\vec{r},t \xrightarrow[]{} \infty)$ (where $\rhoee$ is the excited state fraction), and thus
\begin{equation}
    \avg{\vec{F}_{\mathrm{rad}}} = \hbar \Gamma \rhoee (\vec{r},t \xrightarrow[]{} \infty) \nabla \phi(\vec{r}).
\end{equation}
The radiation pressure force has a simple interpretation in the case of a single beam or the case of balanced counterpropagating beams. For a single plane-wave beam, $\nabla \phi(\vec{r}) = \vec{k}$, such that the force is exactly the momentum kick of a single photon $\hbar \vec{k}$, multiplied by the scattering rate $ \Gamma \rhoee$. In the case of a standing wave, $\nabla \phi (\vec{r})=0$, leading to a zero mean force.

In an experimentally realistic situation, two counterpropagating light beams in one dimension have different intensities, such that the spatially dependent Rabi frequency is
\begin{equation}
\label{eq:omega_cp}
    \Omega(x) = \Omega_1 e^{-i k x} + \Omega_2 e^{i kx},
\end{equation}
where $k=\abs{\vec{k}}$. For this light field, the phase $\phi(x)$ is
\begin{equation}
    \tan [\phi(x)]  = \tan(kx) \frac{\Omega_1-\Omega_2}{\Omega_1+\Omega_2},
\end{equation}
which leads to 
\begin{equation}
    \nabla \phi(x)  = \frac{k(\Omega_1^2-\Omega_2^2)}{\Omega_1^2+\Omega_2^2+2\Omega_1\Omega_2 \cos(2 k x)}.
\end{equation}

\begin{figure*}
    \centering
    \includegraphics[width=\textwidth]{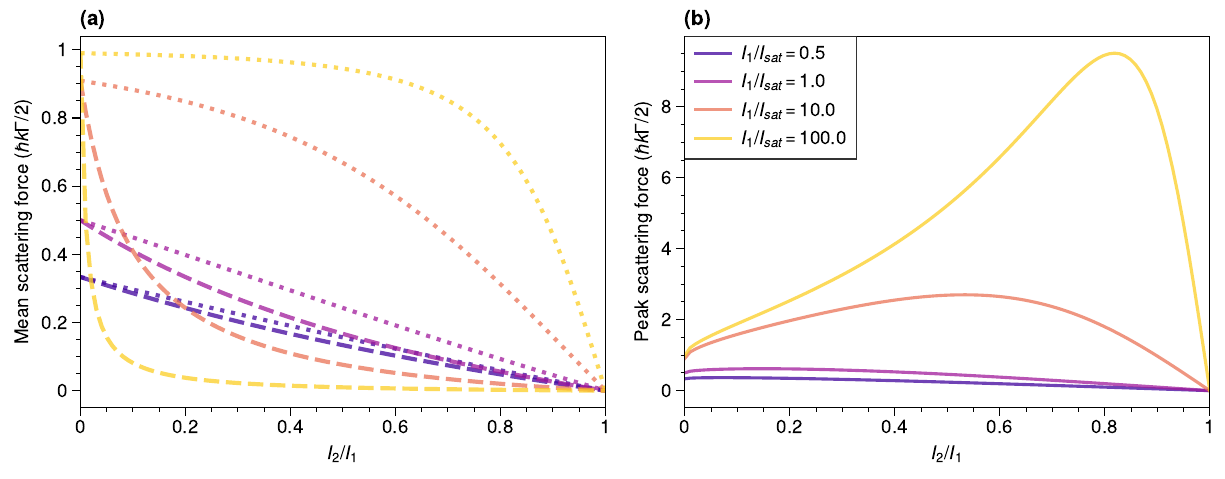}
    \caption{\textbf{Spatially averaged and non-averaged mean radiation pressure force.} (a) Spatially averaged mean radiation pressure force at resonance experienced by an atom in an imbalanced standing wave. Different colors represent different intensities of the first beam $I_1/\Isat$. The dashed lines represent the force when assuming that the two beams are independent and are approximately the force experienced by alternating the two counterpropagating beams. The dotted lines represent the force when two beams interfere if the beams are not pulsed alternatingly. (b) The largest mean radiation pressure force at the trough of the standing wave. When $I_1\gg I_\text{sat}$, the peak radiation pressure force can be much larger than $\hbar k \Gamma/2$.
    }
    \label{fig:mean_peak_force}
\end{figure*}

The expression for the mean force can be simplified by noting that $I_j/\Isat=2\Omega_j^2/\Gamma^2$  and defining $i_j=I_j/\Isat$. Then we can write 
\begin{equation}
    \nabla \phi(x)  = \frac{k(i_1-i_2)}{i_1+i_2+2\sqrt{i_1 i_2}\cos(2kx)}.
\end{equation}
For $\Delta=0$, $s(x)=i_1+i_2+2\sqrt{i_1 i_2}\cos(2kx)$ and thus
\begin{equation}
    \nabla \phi(x)  = \frac{k(i_1-i_2)}{s(x, \Delta=0)}.
\end{equation}

Finally, 
\begin{equation}
    \avg{\vec{F}_{\mathrm{rad}}} = \frac{\hbar k \Gamma}{2} \frac{i_1-i_2}{1+s(x, \Delta=0)}.
\end{equation}
Integrating this expression in space, we get the spatially averaged mean radiation pressure force
\begin{equation}
\label{eq:mean_rad_force}
    \overline{\avg{\vec{F}_{\mathrm{rad}}}} = \frac{\hbar k \Gamma}{2} \frac{i_1-i_2}{\sqrt{(i_1+1)^2+(i_2+1)^2-2 i_1 i_2 - 1}}.
\end{equation}
Note that the equivalent expression for two independent (non-alternating) beams would be
\begin{equation}
    \overline{\avg{\vec{F}_{\mathrm{rad}}}}_{\mathrm{ind}} = \frac{\hbar k \Gamma}{2} \frac{i_1-i_2}{(i_1+1)(i_2+1)}.
\end{equation}

Figure~\ref{fig:spatial_scattering_force} presents the spatial dependence of the various quantities related to atoms interacting with imaging light. Figure~\ref{fig:spatial_scattering_force}(a) shows the total Rabi frequency $\Omega(x)$, given by Eq.~\ref{eq:omega_cp} for $\Omega_1=2.4 \Gamma$ and $\Omega_2=2.1 \Gamma$. The line's color represents the phase of the standing wave, showing a rapid change close to the minima of the Rabi frequency. The saturation parameter $s(x)$ and the excited state fraction $\rhoee(x)$ are shown in Figures~\ref{fig:spatial_scattering_force}(b) and \ref{fig:spatial_scattering_force}(c), respectively. The orange and green lines show the corresponding parameters for the individual beams. Figure~\ref{fig:spatial_scattering_force}(d) shows the spatial dependence of the radiation pressure force in units of $\hbar k \Gamma/2$. The force develops peaks close to the nodes of the standing wave due to the rapid variation of phase $\phi(x)$, even though $\rhoee(x)$ is small there. This situation is close to the situation simulated in Figure~\ref{fig: Gaussian_SSH}(b) and Figure~\ref{fig: Gaussian_SSH}(c), where the atom is simulated to be in the antinode of imaging light.

Figure~\ref{fig:mean_peak_force}(a) shows the results of the expression Eq.~\ref{eq:mean_rad_force} for varying intensity of the first beam $I_1$, denoted by different colored curves and different ratios between the intensities $I_2/I_1$. The independent beam prediction is plotted with dashed lines. For $I_1/\Isat \ll 1$, the two beams can be considered independent even when continuously on, and the exact value of the force matches the force from two independent beams. For larger intensities, the mean radiation pressure force can quickly rise with increasing imbalance between the beams, leading to a larger radiation pressure force than the naive expectation from two independent beams. Note that the mean force is never larger than the maximum radiation pressure force from a single beam: $\hbar k \Gamma/2$. In our experiment, we typically work at $I_1/\Isat \approx 20$. Since we have a 5\% intensity imbalance between the beams from shot to shot, the atoms experience a force equivalent to being pushed by roughly $0.2\hbar k\Gamma/2$. The largest mean radiation pressure force experienced by the atoms is plotted in Figure~\ref{fig:mean_peak_force}(b), close to the nodes of the imbalanced standing wave.

\subsubsection{Experimental implications of high-saturation imaging}
\label{experimental_implications}

In contrast, if only one of the highly saturated beams is on at a time, fluctuations of the beam power do not significantly affect the overall scattering rate and momentum kick. At $I_1/\Isat=20$, if $I_2/I_1$ fluctuates by 50\%, the mean radiation pressure force barely changes by a few percent when the beams are alternatingly pulsed. However, if both beams are on at the same time, the atoms almost only see momentum kicks from one of the beams and can be accelerated away very quickly, resulting in low imaging fidelity. Experimentally, due to intensity disorder over the imaging beams as well as fiber coupling efficiency fluctuations, we systematically observe streaking effects on the camera whose direction varies from shot to shot and even from site to site in the same shot.

We pulse the two imaging beams alternatingly with a pulse length of \qty{400}{ns} (Figure~\ref{fig: SSH_chain}(f)). The pulse length is chosen to not be too short such that the finite rise and fall time of the Acousto-Optic Modulators (AOMs) dominate the pulse times, and not too long such that the net momentum kicks from the pulses push the atoms too far relative to the spacing of the accordion lattices. $I_2$ is pulsed at half duration at the beginning and the end of the pulse sequence to reduce the movement of the atoms away from the original position. We set our RF sources to generate pulses, and then fine-tune the positions of the AOMs and the RF cable lengths to match the rising and falling edges of the two beams to within a few nanoseconds. Since AOMs have wide clear apertures, there can be significant differences in the delay time between the onset of an RF pulse and the onset of the laser beam depending on the distance from the transducer on the AOM to the laser beam. Therefore, simply using an RF switch to alternate the RF power between two AOMs does not guarantee that the two beams will not overlap. Therefore, we use two separate RF sources and monitor the photodiode signal to make sure there is no overlap between the beam pulses on the atoms.

The effects derived above need the two beams to interfere. Now we discuss the case when the polarization of the two beams is orthogonal and both beams are always on. In this case, the mechanism derived above will no longer come into effect. However, the populations are dependent on the $\Omega$ of the two beams and the momentum kick probability will be strongly influenced by the $m_J$ state of the atoms. (For erbium $J=6$) $m_J$ will then stochastically change during the imaging procedure, even if we set the beams to the same intensity. Therefore, a configuration with non-pulsed beams of orthogonal polarization likely does not perform better than the alternatingly pulsed beams demonstrated in this work.

We tried to introduce relative detuning to the counter-propagating beams to make a running wave instead of a standing wave but did not see significant improvement in the histogram.

\subsection{Free space imaging simulation using alternating pulsed beams}
\label{simulation}

First, we simulate how an individual atom would scatter light as it is being kicked by the photons it absorbs and spontaneously emits. We evolve the system on a time step of \qty{0.1}{\ns}, which is two orders of magnitude smaller than the excited state lifetime. We keep track of the $m_F$ state of the atom, which gives rise to different transition probabilities due to different Clebsch-Gordan coefficients. At each time step, we use a random number generator to decide whether the atom will spontaneously emit a photon. We use the rate equation to track the population (SM~\ref{three_level_system}). The direction of the emitted photon is also sampled from a distribution taking into account the polarization of the light and the atom quantization axis (SM~\ref{collection_efficiency}). The atom velocity is updated in response to the momentum kick from the photon. We also simulate the termination of fluorescence with the estimated branching ratio of $4.5\times10^{-5}$. Once the atoms branch into dark states, we assume they no longer scatter photons, since the dark state lifetime is typically much longer than the total imaging duration of a few microseconds. After evolving for the total imaging duration, we extract a list of photon scattering events that record the atom position and the photon direction. We repeat this simulation tens of thousands of times and create a pool to sample from in the following steps. The simulation tracks atom movement in all three dimensions because the low depth of focus of our high NA objective (SM~\ref{objective}) can significantly change the point spread function as atoms move out of the plane.

Second, we arrange a randomly filled square lattice with either zero or one atom per site. We then sample each atom's scattering distribution from the pool generated above. For each photon, we simulate whether it will be detected by the camera sensor considering the optical path transmission of 60 percent and the camera quantum efficiency of 60 percent. Then for each photon, we know the center of its point spread function on the camera by projecting it back on the focal plane. We then convolve the photon position with a point spread function that takes into account the optical wavefront error as well as the uncertainty in the initial atom position. Then we discretize the image into squares corresponding to the pixel size in the image plane. We then account for the EM process by sampling from a Gamma distribution with the appropriate gain.

Third, we take into account the EM CCD camera (Andor iXon Ultra 897 EXF) dark frame performance, including parallel Clock-Induced Charges (CIC), serial CIC, readout noise, and any other imperfections (SM~\ref{camera}). We measure this noise by taking images when there are no atoms present. We then add this background camera noise to the electron counts generated from the second step.

\subsection{Accordion lattice}
\label{accordion}

To dynamically change the accordion lattice spacing while not changing the phase at the center of the lattice, we designed a pair of dove prisms and glued them in-house to a precision of within a micrometer. We purchased commercially available uncoated dove prisms (Thorlabs PS993) and UV-curing optical adhesive (NOA61). We coated one of the dove prisms with broadband 50:50 beam-splitting coating designed to match the refractive index of the optical adhesive. We then designed and machined flexure mount stages to align and glue the two dove prisms together. To change the spacing of the accordion lattice, we use a periscope mounted on a galvanometer before the dove prism pair. The periscope mount is constructed with carbon fiber to reduce the moment of inertia and increase the galvanometer bandwidth. To control the phase of the accordion lattice, we let one of the beams go through an AR-coated 1-mm thick window (phase plate) mounted on another galvanometer. As the galvanometer is rotated, the beam path length changes. More details about the physical setup and alignment procedures are elaborated in \cite{Hebert2021}.

We use a single-pass-doubling ytterbium-fiber-amplified laser (Azurlight Systems), whose wavelength is \qty{488.1}{\nm} and maximum power is \qty{2}{\W}. The beam size in the dove prism pair is around \qty{100}{\um} and the peak intensity is around 25 W/mm$^2$. During testing, we observed slow degradation of the transmission through the dove prism pair when continuously sending in a beam at this level of intensity. Therefore, we mounted the prism pair on a translation stage such that we could move the prism orthogonal to the beam propagation direction if the transmission degrades too much at a particular spot. We have been operating for three years and have not noticed significant degradation yet, since we only need high power for a fraction of a second per experimental shot during imaging.

\begin{figure*}
    \centering
    \includegraphics[width=\textwidth]{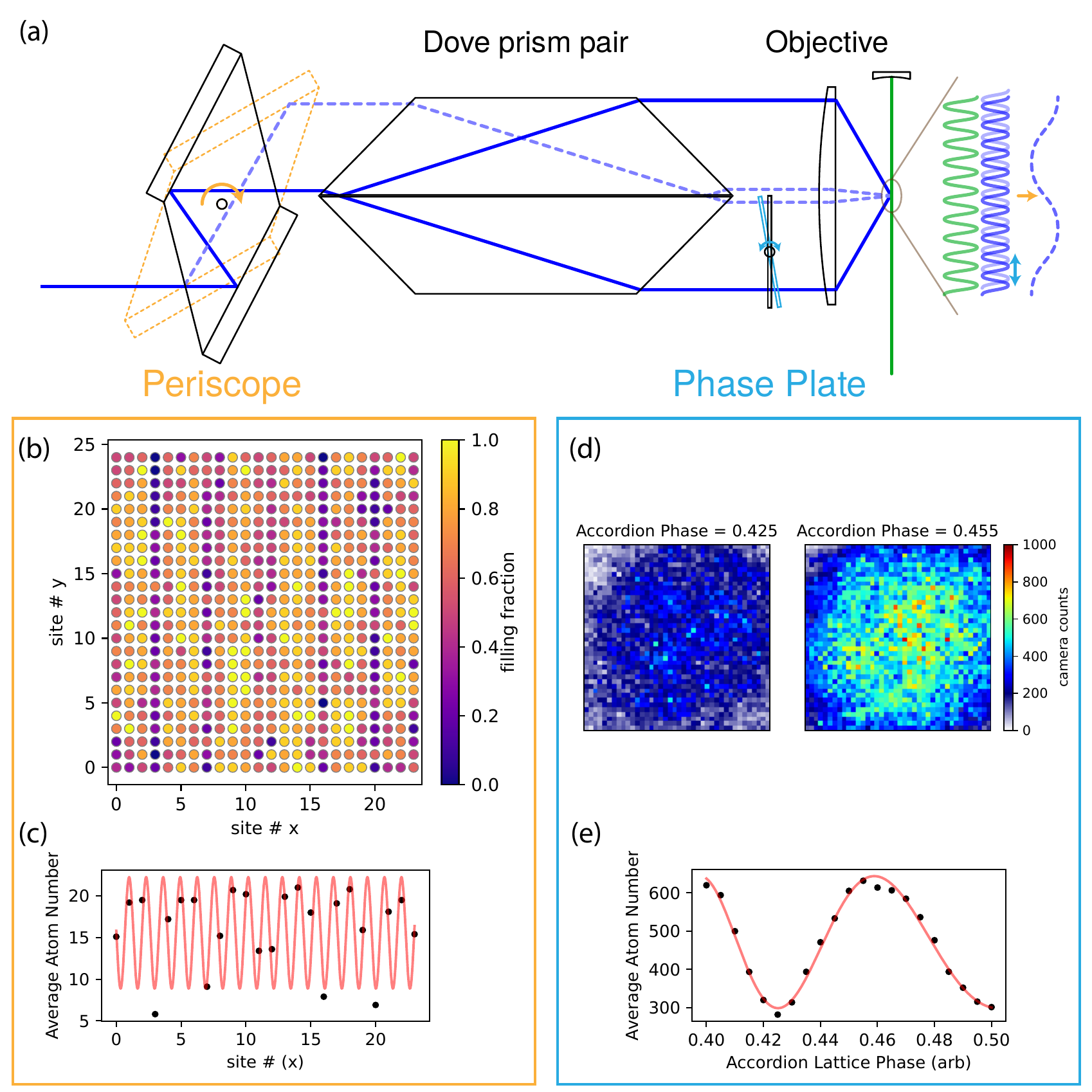}
    \caption{\textbf{Accordion lattice setup and calibration.} The accordion lattice consists of two moving parts: a periscope that controls the lattice spacing and a phase plate that controls the lattice phase. As the periscope is rotated clockwise following the orange arrow in (a), the incident beam position changes to the dashed blue line and the spacing between the exit beams becomes small. Upon going through the objective, the lattice spacing becomes larger. We can calibrate the spacing of the accordion lattice relative to the green retro-reflected lattice in (a) using Moir\'e patterns. An example site-resolved average filling is shown in (b). By summing over the y-axis, we obtain the atom number and fit it with a sinusoidal that matches the accordion lattice spacing in (c). As the phase plate is rotated following the cyan arrow in (a), the accordion lattice phase can be tuned. We calibrate the phase of the accordion lattice relative to the green retro-reflected lattice by quenching off the retro-reflected lattice and quenching on the accordion lattice. After holding and letting the higher band atoms move out of the region of interest, we can take images as shown in (d) as we set the phase to different values, and then fit a sinusoidal pattern to the atom number.
    }
    \label{fig: accordion}
\end{figure*}

To achieve the best transfer fidelity between lattices we must match each lattice's spacing and phase to be commensurate. We conducted an initial calibration of the spacing of the accordion lattices when they were set up for the first time. We loaded a large cloud of atoms into a deep \qty{266}{\nm} retro-reflected tightly spaced lattice with a shallow accordion superlattice superimposed. We then fit the Moir\'e patterns formed by the two lattices to determine the spacing. While spacing calibration only needs to be done once, we calibrate the lattice phase daily. To do so, we load atoms into the tightly spaced lattice and simultaneously quench off the tightly spaced lattice and quench on the accordion lattice within a microsecond. This quench is faster than the trap frequency of roughly 30 kHz. We then hold the accordion for a few hundred milliseconds, during which atoms occupying higher lattice bands due to imperfect phase matching will tunnel away from the region of interest. We finally take a picture (without expanding the accordion lattice for site-resolved imaging) and record the total atom number. We repeat this process at different phases and identify the phase resulting in the largest preserved atom number as the commensurate phase in which both lattices line up (Figure~\ref{fig: accordion}(d) and (e)). The phase does not drift after an experimental warm-up period of a few hours. We cover the accordion lattice optical breadboard with an anodized aluminum enclosure to reduce the disturbance of the optical path length from air currents and maintain a stable temperature.

We study the atom lifetime in the accordion lattice and explore possible heating sources. We measure the lifetime of atoms without the accordion lattice to be \qty{90}{s}. Once we turn on the accordion lattice at a fixed spacing, we observe a lifetime between \qty{10}{s} and \qty{30}{s} depending on the accordion lattice power and spacing. We keep the lattice spacing the same and vary the power. We observe a longer atom lifetime with shallower lattice depth, implying that the loss could be induced by scattering. We also observe a decrease in the lifetime when the spacing is larger than \qty{5}{\um}, possibly related to the decrease of trap frequency when the accordion spacing is expanded too much.

\begin{figure}
    \centering
    \includegraphics[width=0.48\textwidth]{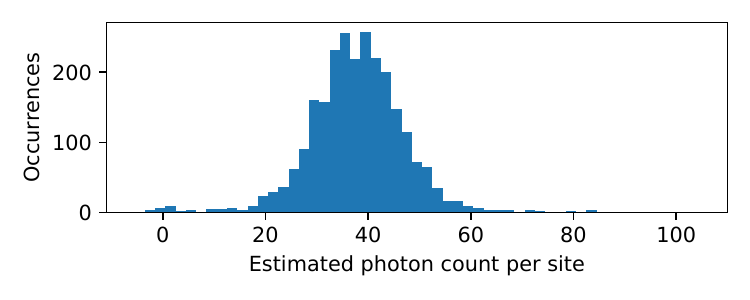}
    \caption{\textbf{Unity-filled Mott Insulator histogram} We look at the center 2 by 2 sites of a unity-filled Mott Insulator of more than 600 shots without post-selection to estimate an upper bound of the loss in our imaging technique to be roughly 1\%.
    }
    \label{fig:transfer_fidelity}
\end{figure}

We measured the central filling of a unity-filled Mott Insulator to be $99.00\pm0.16\%$, which puts an upper bound of the atom loss to 1\% (Figure~\ref{fig:transfer_fidelity}). Since the atoms spend roughly \qty{80}{ms} in the accordion lattices, based on the lifetime measured, we expect on the order of 0.7\% loss directly coming from heating in the accordion lattices without accounting for transferring and expanding. In addition, the imaging fidelity is still affected by non-cycling dark states (the branching ratio is measured in the main text) and has a false negative rate of roughly 0.3\%. This implies that we see very little loss due to the dynamical transfer to and the expansion of the accordion lattices, which is consistent with simulation. Thus, our work establishes the use of accordion lattices as an almost lossless tool for imaging.

We cannot make a similar reliable estimate for the loss probability for $n\geq2$ in the accordion lattices, since we observe non-negligible loss of doublons in our tightly spaced retro-reflecting lattice even before transferring to the accordion lattice, which limits the original fidelity of the Mott insulator with more than one particle per site. We observe that the doublon loss rate depends on the magnetic field and worsens near Feshbach resonances.

We perform adiabatic ramps in the retro-reflected lattices because they have much lower intensity disorder and allow us to achieve regimes where the dipolar interaction is the dominant energy scale \cite{Su2023}. In contrast, the projected accordion lattices have significantly more disorder, making them unfit for adiabatic ramps. However, this disorder does not affect experimental results since we project our quantum states to the number basis after quickly quenching the tunneling in the retro-reflected lattices.

\subsection{In-vacuum objective}
\label{objective}

To project the accordion lattice of \qty{488}{nm} light to match the \qty{266}{nm} spacing of the retro-reflected green lattice, we need a high NA of 0.92. If we use a shorter wavelength for the accordion lattice, we can reduce the required NA of the objective. In addition, the objective does not have to be diffraction-limited for accordion lattice projection. So an ultra-high NA objective is not necessary for this imaging method with accordion lattices. On the other hand, for imaging and potential control, we ideally want the objective to have a high diffraction limited performance to the highest possible NA. Large NA diffraction limited performance allows us to maximize the number of photons we can collect during our short image duration and maximize the resolution of projected arbitrary potentials via Digital Micro-mirror Devices (DMD). We custom-ordered our objective from Special Optics. Due to other experimental requirements (see \cite{Su2023}), we requested the objective to have a hole in the center of each optical element that makes up the objective. We tested that the wavefront error up to an NA of 0.85 is roughly 0.1 waves for our imaging beam at \qty{401}{nm}. In addition, the wavefront error up to an NA of 0.9 is roughly 0.07 waves for our DMD projection beam at \qty{532}{nm}. The test method and results are described in \cite{Krahn2021}.

Such high NA requirements make it desirable to mount the objective inside the vacuum chamber. We verify that the same optical performance can be achieved after baking, a procedure necessary to achieve pressure lower than $10^{-11}$ Torr, which is required for many-body experiments \cite{Su2023}. We test-baked the objective in a separate Ultra-High Vacuum (UHV) chamber at the desired temperature of \qty{95}{\celsius} for 1 week. We removed the objective from the test chamber and measured the wavefront error again to confirm that the objective had not misaligned from the heating and cooling cycle. In addition, we verified that the objective is UHV compatible by checking the final pressure level after the test bake and monitoring the chamber contents using a residual gas analyzer to track the gas component during and after the bake. We then mounted the objective into the experiment chamber and baked it for 4 weeks at \qty{95}{\celsius}. We finally achieved a vacuum lifetime of more than \qty{90}{s} and the pressure has been stable over the past three years without the need to undergo additional rounds of titanium sublimation pumping.

Higher transmission of the scattered light through the imaging optical system and better camera quantum efficiency can help achieve better imaging. The transmission of our imaging beam at \qty{401}{nm} through the objective is measured to be less than 70\%. One reason for the low transmission is that the glass used in the objective absorbs a significant amount of light close to the ultraviolet spectrum. In addition, the anti-reflective coating on each lens is coated for multiple wavelengths and the angle of incidence in this high-NA objective is relatively broad, further reducing the transmission. After the beam goes through the vacuum viewport, multiple dichroic mirrors, lenses, and mirrors, we estimate a total transmission of 60\%. To optimize for the point spread function and crop out the regions where the objective is not diffraction-limited, we apply an iris after the collimated beam exiting the vacuum chamber. Photons are collected over half-angles ranging from $\arcsin(0.3)\approx\ang{17}$ to $\arcsin(0.85)\approx\ang{58}$. We set the quantization axis of our atoms and the polarization of the laser beam to maximize the collection efficiency in SM~\ref{collection_efficiency}.

\subsection{Quantization axis and collection efficiency through the objective}
\label{collection_efficiency}

\begin{figure*}
    \centering
    \includegraphics[width=\textwidth]{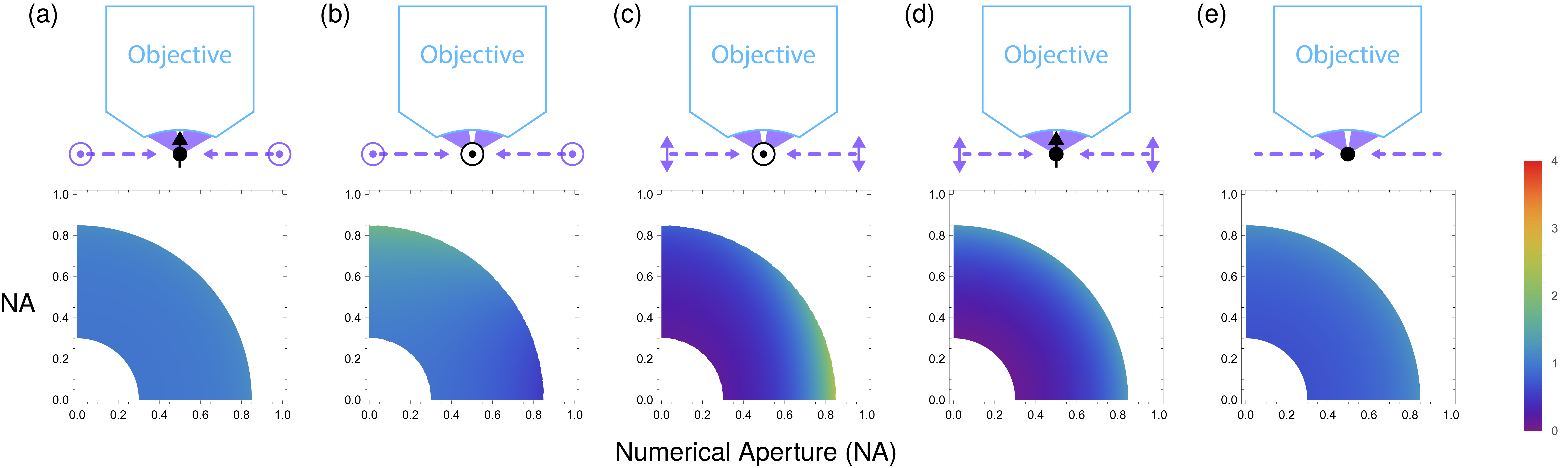}
    \caption{\textbf{Fourier plane image with different polarization.} As we change the light polarization (purple arrow) and the atom quantization axis (black arrow), the calculated Fourier plane image (bottom) as well as the collection efficiency of the objective changes. The collection efficiency with our objective NA from 0.3 to 0.85 are 25.0\% (a), 25.0\% (b), 
    19.5\% (c), 14.0\% (d), and 21.4\% (e).}
    \label{fig:collection_efficiency}
\end{figure*}

The light's polarization and the atom's quantization axis affect our objective's collection efficiency. We define collection efficiency as the ratio of photons collected by the objective over the total number of scattered photons. Here we present the calculations we made to choose the polarization and quantization axis.

\subsubsection{Angular distribution}

The angular distribution of scattered photons is $dw\propto[d_{q,\mu}^1(\theta)]^2(\sin\theta d\theta d\phi)$ \cite{Littlejohn2020}. $d_{q,\mu}^1(\theta)$ is D-matrix in Euler angle form, $q$ is the change of $m$ state of the atom, and $\mu$ is the detected photon polarization. We calculate different $q$ cases separately because they have different Clebsch–Gordan coefficients and therefore different scattering probabilities. For simplicity, we take the basis of circularly polarized light, where we only consider the orthogonal basis of $\mu=\pm1$.

when $q=1$

\begin{align*}
    d_{1,1}^1&=\frac{1}{2}(1+\cos\theta)\\
    d_{1,-1}^1&=\frac{1}{2}(1-\cos\theta)\\
\end{align*}

when $q=0$

\begin{align*}
    d_{0,1}^1&=\frac{1}{\sqrt{2}}(\sin\theta)\\
    d_{0,-1}^1&=-\frac{1}{\sqrt{2}}(\sin\theta)\\
\end{align*}

when $q=-1$

\begin{align*}
    d_{-1,1}^1&=\frac{1}{2}(1-\cos\theta)\\
    d_{-1,-1}^1&=\frac{1}{2}(1+\cos\theta)\\
\end{align*}

\subsubsection{Collection efficiency}
In our objective, we collect light of both $\mu=\pm1$ polarization. So we sum up their squares.

when $q=\pm1$, $dw\propto(3+\cos2\theta)(\sin\theta d\theta d\phi)$

when $q=0$, $dw\propto\sin^2\theta(\sin\theta d\theta d\phi)$

We present two cases.

In the first case, we consider the objective axis aligned to the atom polarization axis. When $q=\pm1$, collection efficiency (without normalization) is 

$$\int^{\theta\max}_{\theta\min} d\theta \int^{2\pi}_0d\phi(3+\cos2\theta)(\sin\theta)$$

With our objective NA from 0.3 to 0.85, we get a collection efficiency of 25\% (Figure~\ref{fig:collection_efficiency}(a)).
When $q=0$, collection efficiency (without normalization) is 

$$\int^{\theta\max}_{\theta\min} d\theta \int^{2\pi}_0d\phi\sin^2\theta(\sin\theta)$$

With our objective NA, the collection efficiency is 14\%(Figure~\ref{fig:collection_efficiency}(d)).

In the second case, we consider the objective axis orthogonal to the atom polarization axis. We perform a change of variable.
\begin{align*}
\begin{aligned}
    \theta'&=\arccos(\sin\phi\sin\theta)\\
    \phi'&=\arctan(\cos\phi\tan\theta)\\
    \theta&=\arccos(\cos\phi'\sin\theta')\\
    \phi&=\arctan(1/(\sin\phi'\tan\theta'))\\
\end{aligned}
\end{align*}
For simplicity, we define $g(u,v)=\arccos(\sin u\cos v)$

In general, we have

\begin{align*}
\begin{aligned}
w&\propto \int f(\theta) d\Omega\\
&=\int d\phi\int d\theta f(\theta)\sin\theta\\
&=\int d\phi' \int d\theta' f(g(\theta',\phi'))\abs{\frac{\partial(\theta,\phi)}{\partial(\theta',\phi')}}\sin(g(\theta',\phi'))
\end{aligned}
\end{align*}

where $\abs{\frac{\partial(\theta,\phi)}{\partial(\theta',\phi')}}=\abs{\frac{\partial\theta}{\partial\theta'}\frac{\partial\phi}{\partial\phi'}-\frac{\partial\theta}{\partial\phi'}\frac{\partial\phi}{\partial\theta'}}$ is the Jacobian determinant.

When $q=\pm1$, $f(\theta)=(3+\cos2\theta)$. Plugging in our objective clear aperture, we get a collection efficiency of 19.5\% (Figure~\ref{fig:collection_efficiency}(c)). When $q=0$, $f(\theta)=(\sin\theta)^2$, and the collection efficiency is 25\% (Figure~\ref{fig:collection_efficiency}(b)).

Assuming the atom is unpolarized, $f(\theta)=1$. Plugging in our objective clear aperture, we obtain a collection efficiency of 21.4\% (Figure~\ref{fig:collection_efficiency}(e)).

\subsubsection{Fourier plane photon density}
The power density in the Fourier plane depends on the quantization axis and light polarization. With the wavefront error from the objective, one can optimize the point spread function by choosing different Fourier plane power densities. We chose the configuration of Figure~\ref{fig:collection_efficiency}(a) because it gives the best collection efficiency and has an almost uniform Fourier plane photon density, resulting in the best point spread function among the options shown here.

When the objective optical axis is aligned with the atom quantization axis, we map the $\theta$ variable to $r$ (NA). We know that $r=\sin\theta$.

For $q=\pm1$, we have $dw\propto(3+\cos2\theta)(\sin\theta d\theta d\phi)$. After normalization, we get $dw=\frac{3}{8\pi}\frac{2-r^2}{\sqrt{1-r^2}}(r dr d\phi)$.

For $q=0$, we have $dw\propto(\sin\theta)^2(\sin\theta d\theta d\phi)$. After normalization, we get $dw=\frac{3}{4\pi}\frac{r^2}{\sqrt{1-r^2}}(r dr d\phi)$.

We have $\frac{dw}{rdrd\phi}=\frac{r^2}{\sqrt{1-r^2}}$

When the objective optical axis is orthogonal to the atom quantization axis, we change $\theta'$ to $r=\sin\theta'$

\begin{align*}
\begin{aligned}
    r&=\sin(\arccos(\sin\phi\sin\theta))\\
    \phi'&=\arctan(\cos\phi\tan\theta)\\
    \theta&=\arccos(r\cos\phi')\\
    \phi&=\arctan(1/(\sin\phi'\tan(\arcsin(r))))\\
\end{aligned}
\end{align*}

In general, we have
$$dw\propto f(\arccos(r\cos\phi'))\abs{\frac{\partial(\theta,\phi)}{\partial(r,\phi')}}\sin(\arccos(r\cos\phi'))drd\phi'$$

Where $\abs{\frac{\partial(\theta,\phi)}{\partial(r,\phi')}}=\abs{\frac{\partial\theta}{\partial r}\frac{\partial\phi}{\partial\phi'}-\frac{\partial\theta}{\partial\phi'}\frac{\partial\phi}{\partial r}}=\frac{r}{\sqrt{(1-r^2)(1-r^2\cos^2(\phi'))}}$ is the Jacobian determinant.

\subsection{Three level systems}
\label{three_level_system}

In our simulation (SM~\ref{simulation}), we keep track of the $m_J$ state of an atom during its evolution. At specific $m_J=m$, since the light polarization we use can drive both $\sigma_+$ and $\sigma_-$ transitions, we calculate the steady state population of the three-level system between $m_J=m$, $m_J'=m+1$, and $m_J'=m-1$. The steady-state excited population multiplied by $\Gamma dt$ represents the probabilities of projecting to $m_J'$ states. In the simulation (SM~\ref{simulation}), we evolve the atomic state based on these probabilities and account for the different scattering angle distributions for $\sigma_\pm$ and $\pi$ transitions. The first few of the optical Bloch equations under homogeneous broadening are

\begin{align*}
\begin{aligned}
    \partial_t\rho_{33}&=i\frac{\Omega_{13}}{2}(\tilde{\rho}_{31}-\tilde{\rho}_{13})-\Gamma\rho_{33}\\
    \partial_t\rho_{22}&=i\frac{\Omega_{12}}{2}(\tilde{\rho}_{21}-\tilde{\rho}_{12})-\Gamma\rho_{22}\\
    \partial_t\rho_{11}&=\Gamma(\rho_{33}+\rho_{22})-\frac{i}{2}(\Omega_{13}(\tilde{\rho}_{31}-\tilde{\rho}_{13})+\Omega_{23}(\tilde{\rho}_{21}-\tilde{\rho}_{12}))\\
\end{aligned}
\end{align*}

Since we are looking for steady-state solutions with negligible detuning due to Doppler shift (even scattering 1000 photons all in the same direction gives a 15MHz detuning, which is only half of the linewidth of the transition), we can set $\Gamma_{31}=\Gamma_{21}=\Gamma$ and $\Delta=0$ \cite{Sen2014}. We assume that the coherence does not change, such that coherence terms can be eliminated from the above equations. For simplicity, we ignore the coherent transfer to $m_J=m\pm2$ and higher order processes. We then get $\rho_{ii}=\frac{\Omega_{1i}^2}{\Gamma^2+2(\Omega_{12}^2+\Omega_{13}^2)}$.

We shine highly saturated light to reduce the effect of different Clebsch-Gordan coefficients and maximize the scattering rate at any Zeeman sublevel. With a beam size of a few hundred microns, this can be achieved with ten milliwatts of laser power for our broad $\Gamma=2\pi\times30$MHz transition.

\subsection{Definition of terms}
\label{definition}

\subsubsection{Saturation intensity}
Here we clarify our definition of the saturation intensity. As stated in Eq. (5.244) of \cite{steck2007quantum}, we define $I_\text{sat}=\frac{\hbar\omega_0A_{21}}{2\sigma_0}$. As noted in Eq. (3.21) and (3.22), when the atom polarization is random, we get $\sigma(\omega)=\frac{\lambda^2}{2\pi}$, but when the atomic dipole moments are aligned with the field polarization, we get $\sigma(\omega)=\frac{3\lambda^2}{2\pi}$. In our case, we have the atoms polarized, so $I_\text{sat}=\frac{\hbar\omega_0\Gamma}{2\frac{3\lambda^2}{2\pi}}=\frac{\pi\hbar\omega_0\Gamma}{3\lambda^2}$. In contrast, Eq. (5.253) assumes random atom polarization and therefore is a factor of $3$ larger. For our transition, the saturation intensity is \qty{56}{\mW/\cm^2}. Rabi frequency in the unit of excited state lifetime can be calculated with $\Omega=\Gamma\sqrt{\frac{I}{\Isat}/2}$.

\subsubsection{Imaging Fidelity}

We define the imaging fidelity of distinguishing between $n$ and $n+1$ atoms per site as the average of the false positive and false negative rates. The false positive (negative) refers to the rate of the case when the total camera count is above (below) the preset cutoff but there are still $n$ ($n+1$) atoms on the site. The imaging fidelity here does not consider the atom loss during the transfer from the small-spacing lattice to the accordion lattice and the expansion of the accordion lattice.

\subsection{EM CCD spurious count and its effect on binarization performance}
\label{camera}

\begin{figure}
    \centering
    \includegraphics[width=0.48\textwidth]{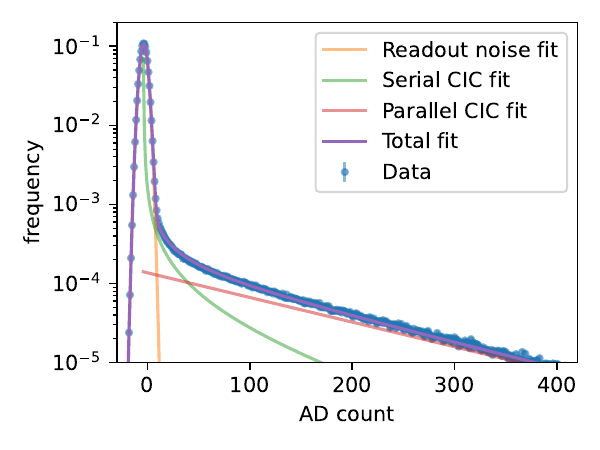}
    \caption{\textbf{Example fit of camera dark frames} We take images without incident light and analyze the histogram. We fit the histogram (blue) with three components: the Gaussian readout noise (yellow), the serial CIC (green), and the parallel CIC (red). The sum of all these three components is shown in purple. We extract the readout noise and the CIC probability from the fit. We plot an example data where we set the readout speed to 1MHz and the vertical shift to 1.7us, which is the combination that gives us the lowest readout noise. 
    }
    \label{fig:dark_frame_hist}
\end{figure}

\begin{table}
\begin{ruledtabular}
Vertical shift \qty{3.3}{\us}
\begin{tabular}{c|c|c|c|c}
    Readout speed (MHz) & 1 & 5 & 10 & 17 \\
    \hline
    Readout noise std (electrons) & 0.025 & 0.068 & 0.099 & 0.083 \\
    Serial CIC (percentage) & 4.7 & 4.5 & 4.0 & 2.5 \\
    Parallel CIC (percentage) & 3.0 & 2.9 & 2.9 & 3.0
\end{tabular}
\end{ruledtabular}
\vspace{1em}
\begin{ruledtabular}
Vertical shift \qty{1.7}{\us}
\begin{tabular}{c|c|c|c|c}
    Readout speed (MHz) & 1 & 5 & 10 & 17 \\
    \hline
    Readout noise std (electrons) & 0.025 & 0.070 & 0.102 & 0.084 \\
    Serial CIC (percentage) & 3.9 & 3.4 & 2.6 & 1.8 \\
    Parallel CIC (percentage) & 2.0 & 1.8 & 1.8 & 1.8
\end{tabular}
\end{ruledtabular}
\caption{\label{tab:dark_frame} The measured performance of our EM CCD camera with the slowest (top) and the second slowest (bottom) vertical shift speed.}
\end{table}

We characterize the dark count of our Andor iXon Ultra 897 EXF. We take dark images with a maximum gain of 1000. To maintain good electron transfer fidelity throughout the whole image, we keep the vertical shift speed at \qty{1.7}{\us} and \qty{3.3}{\us} only, since we observe part of the image will lose signal if we overclock the vertical shift too much. We analyze the histogram by fitting it with the sum of Gaussian readout noise, serial CIC, and parallel CIC (Figure~\ref{fig:dark_frame_hist}). With the parallel CIC fit, we extract the conversion of camera count to photoelectrons and use it with the Gaussian fit to calculate the readout noise. We summarize our results in Table~\ref{tab:dark_frame}. The Clock-Induced Charges broaden the histogram peak with 0 atoms per site and limit our imaging fidelity since we have a finite branching ratio. With qCMOS cameras and proper magnification, we expect to achieve even faster imaging with higher fidelity.

With 512 by 512 pixels, we have enough space to image 4000 sites with 8 by 8 pixels per site using binarization. Cameras with 1024 by 1024 or more pixels are easily available commercially, enabling the imaging of tens of thousands of particles.

\nocite{*}

\bibliography{SingleSiteImaging}

\end{document}